\documentclass{aa}
\usepackage{lmodern}
\usepackage{txfonts}
\usepackage{hyperref}
\hypersetup{
    colorlinks=true,
    linkcolor=blue,
    filecolor=magenta,      
    urlcolor=blue,
    citecolor=blue
}
 
\newcommand{\avg}[1]{\langle #1 \rangle}

\newcommand{\prcb}{P_{\rm RCB}}
\newcommand{\trcb}{T_{\rm RCB}}
\newcommand{\rrcb}{R_{\rm RCB}}
\newcommand{\rhorcb}{\rho_{\rm RCB}}
\newcommand{\rgas}{\mathcal{R}}
\newcommand{\Mc}{M_{\rm c}}
\newcommand{\Rc}{R_{\rm c}}
\newcommand{\Matm}{M_{\rm atm}}
\newcommand{\rbprime}{R_{\rm B}'}
\newcommand{\Mpl}{M_{\rm pl}}
\newcommand{\rgap}{R_{\rm gap}}
\newcommand{\nablaad}{\nabla_{\rm ad}}

\newcommand{\Rpl}{R_{\rm pl}}
\newcommand{\Miso}{M_{\rm iso}}
\newcommand{\difffrac}[2]{\frac{{\rm d} #1}{{\rm d} #2}}

\begin{document}
\title{A primordial radius valley as a consequence of planet formation}
\author{Jesper Nielsen\inst{1}
\and
Anders Johansen \inst{1,2}
\and
Komal Bali\inst{3}
\and
Caroline Dorn\inst{3}
}

\institute{
Center for Star and Planet Formation, Globe Institute, University of Copenhagen, Øster
Voldgade 5-7, 1350 Copenhagen, Denmark email: \href{mailto:jesper.nielsen@sund.ku.dk}{jesper.nielsen@sund.ku.dk}
\and
Lund Observatory, Department of Physics, Lund University, Box 43, 22100 Lund, Sweden
\and
Institute for Particle Physics and Astrophysics, ETH Zürich, Otto-Stern-Weg 5, 8093 Zürich, Switzerland
}

\date{Accepted 14 February 2025}

\abstract{The radius distribution of close-in planets has been observed to have a bimodal distribution with a dearth of planets around $\sim$1.5-2.0 $R_\oplus$ commonly referred to as the ''radius valley''. The origin of the valley is normally attributed to mass-loss process such as photoevaporation or core-powered mass loss. Recent work, however, has suggested that the radius valley may instead arise as a consequence of gas accretion by low-mass planets. In this work we therefore aim to investigate the formation of a primordial radius valley from the formation of planet cores through pebble accretion up until the dissipation of the protoplanetary disc and subsequent contraction of accreted atmospheres. The goal of this work is to explore the conditions for forming a primordial radius valley from first principles of planet formation theory, rather than attempting to explain the detailed structure of the observed valley. We use an analytical model with minimal assumptions to estimate the contraction rate of atmospheres and, indeed, find the formation of a primordial radius valley. The planets smaller than the valley did not reach the pebble isolation mass, which is required for the planets to cool down sufficiently to be able to accrete a significant amount of gas. We also estimate the slopes of the radius gap as a function of orbital period for the intrinsic population as well as for planets with orbital periods $<$100 days. For the intrinsic population, the radius gap follows the pebble isolation mass and increases with increasing orbital period, while for close-in planets the direction of the slope reverses and decreases with increasing orbital period. We find that planets smaller than the radius valley are predominantly rocky while the population of planets larger than the valley consists of a mixture of rocky and water-rich planets.}

\maketitle
\section{Introduction}
The Kepler mission \citep{borucki2010_kepler} has given us a lot of insight into the formation and evolution of planetary systems. One of the key findings from the Kepler mission is the apparent lack of planets with radii in the range $\sim$1.5-2.0 $R_\oplus$, for orbital periods $<$100 days \citep{fulton2017_valley}. This is commonly referred to as the ''radius valley''. Follow-up studies, utilising precise stellar parameters from Gaia \citep{gaia2016} and more precise planet radii measurements confirmed these findings and found that the location of the valley increases with increasing stellar masses and decreases with increasing orbital periods \citep{fultonpetigura2018_gap,berger2020b,Ho_vaneylen2023_valleymstar}. 

Observed planets smaller than $\sim$1.6 $R_\oplus$ have been inferred to have a rocky composition. In contrast, larger planets have too low densities to be composed mainly of metals and silicates, which implies that larger planets require an atmosphere of low mean molecular weight to explain their masses and sizes \citep{rogers2015_density,wolfganglopez2015_rocky}. Further, only a small atmosphere mass fraction ($\sim$1-3\%) is required in order to inflate the radius significantly, thereby doubling the size of the planet \citep{lopezfortney2014_inflate,owen_wu2017_PE}. This indicates that planets smaller than the radius valley have very light to no atmospheres, while planets larger than the radius valley host an atmosphere with a mass fraction of at least $\sim$1\%.

Attempts to explain the radius valley have therefore focused on trying to understand how planets lose their atmospheres, populating the smaller end of the radius valley. There have been two major mechanisms that both shape the distribution of planetary radii during their evolution: photoevaporation \citep{lopezfortney2013_evap,owenwu2013_evap,owenwu2016_boiloff,owen_wu2017_PE} and core-powered mass loss \citep{ginzburg2018_cpm,gupta2019_cpm,gupta2020_cpm,gupta2021_cpm,gupta2022_cpm}. After the pressure support of the disc is removed, gas at the Bondi radius will escape at the Bondi rate. In order to sustain mass-loss, the remaining gas needs to expand out towards the Bondi radius. Such an expansion requires a supply of heat. The mass loss in the photoevaporation model is mainly driven from high-energy flux from the star, accelerating the atoms in the atmosphere, which causes heating, expansion and an outflow. Core-powered mass loss is instead mainly driven by the remaining thermal energy of the planet, combined with the bolometric illumination by the central star \citep{ginzburg2016_cpm,ginzburg2018_cpm}. The position of the radius gap has been observed to decrease with increasing orbital period \citep{fulton2017_valley,petigura2022_rgapvstar}, which both mass loss models are able to predict \citep{owen_wu2017_PE,gupta2019_cpm}. However, \citet{cloutiermenou2020_mdwarfslope} found that the radius gap increases with increasing orbital period for M-dwarfs, which might hint towards a separate planet formation process for planets around these types of stars as an increasing radius gap with increasing orbital period is generally predicted by planet formation after the dissipation of the protoplanetary disc \citep{lopezrice2018_gaspoor}.

The two presented mass loss models nevertheless predict different dependencies on the stellar mass. As the core-powered mass loss mechanism is driven by the planet's internal energy as well as the bolometric luminosity, it predicts that given a fixed incident bolometric flux onto the planet, there is no dependence on the luminosity (and hence the mass) of the host star. The photoevaporation model, in contrast, predicts a stellar mass dependency of the photoevaporation timescale since the high-energy luminosity fraction does scale with stellar mass \citep{jackson2012_xray,shkolnik2014_xray,rogers2021_compare}.  \citet{fultonpetigura2018_gap} investigated the stellar mass dependencies of the radius valley and found results more consistent with the photoevaporation model. \citet{berger2023_cpmobs} used data from Gaia DR3 \citep{gaiadr3} and found that the predictions for the observed stellar mass scaling given a fixed flux from both models was within the observed uncertainties. In contrast, the observed flux scaling given a fixed stellar mass could only be explained by the core-powered mass loss model. Further, \citet{loyd2020_cpmobs} found that the observed radius gap did not vary with stellar mass after fixing incident fluxes, indicating that photoevaporation is not favoured over core-powered mass loss. Core-powered mass loss also seems favoured when considering the ages of sub-Neptune host stars. Observations indicate that the occurrence rate of sub-Neptunes of similar age as the timescale for photoevaporation ($\sim$a few 100 Myr) is larger than the occurrence rate of older sub-Neptunes of similar age as the core-powered mass loss timescale ($\sim$a few Gyr) \citep{berger2020b,christiansen2023_k2ages}. These results could indicate that sub-Neptunes lose their atmospheres on timescales similar to that of the core-powered mass loss mechanism. However, recent work by \citet{owen_schlichting2024_cpmpecombo} has shown that both core-powered mass loss and photoevaporation can contribute to the loss of atmospheres after disc dispersal and that they are dominant in different regions of the parameter space. Core-powered mass loss is found to be mostly dominant for planets with low core masses and high atmosphere mass fractions while photoevaporation dominates for more massive planets with lower atmosphere mass fractions. This means that a planet can initially lose mass through the core-powered mass loss mechanism and move into a region where photoevaporation dominates. A planet that is initially experiencing photoevaporative mass loss will however not transition into core-powered mass loss \citep{owen_schlichting2024_cpmpecombo}. This means that it might be difficult to disentangle the two mechanisms as they are both expected to contribute to the mass loss of planetary atmospheres.

It has also been shown that the radius gap can be reproduced without the need for a gaseous envelope by instead allowing for different bulk compositions of the planets. The two peaks in the radius distribution can then emerge by simply considering the presence of water in the form of vapour on the planet. Water-rich planets would naturally be inflated compared to rocky cores, creating the two peaks in the radius distribution \citep{venturini2020_water,burn2024_waterworlds}. Further, after disc dissipation, multiple planets may be locked in resonant chains which become unstable, causing orbital crossings and potentially collisions between planets. Such collisions would strip away any accreted atmospheres and only bare cores would remain \citep{liu2015_giantimpactstrip}. Dense, rocky cores would then constitute the peak at $\sim$1.5 $R_\oplus$ while water-rich cores ($\gtrsim$50\% water mass fraction) of similar masses would be larger and constitute the peak at $\sim$2.1 $R_\oplus$ on the other side of the valley \citep{izidoro2021_breakingchains,izidoro2022_migvalley}. Further, the radius valley has been found to be significantly shallower around low-mass host stars. This trend could be explained by water-rich planets with lower densities and therefore inflated core radii \citep{ho2024_valleyslope}.

The emergence of a radius valley has also been explained as a primordial feature without the need for mass loss. Instead the peak at smaller radii consists of planets that are not able to accrete enough gas to become inflated to $\sim$$2\,R_\oplus$. For a given core mass, there exists a maximum atmosphere mass that can be accreted before it becomes fully isothermal with the same temperature of the disc. For small cores ($\sim$1-2$\,M_\oplus$), this mass is small enough to not grow beyond the radius valley \citep{leeconnors2021_primordial,lee2022_gapwithoutmassloss}. 

Finally, the radius valley can also be explained in the context of core formation. Planets that form close to the star are generally thought to form through a combination of pebble accretion and giant impacts \citep{lambrechts2019_gpform,drazkowska2023_pfreview}. While the protoplanet is still embedded within a protoplanetary disc, the accretion of pebbles will heat up the planet enough to prevent the contraction of a significant primordial atmosphere. This will prevent the accretion of any significant amount of gas \citep{lambrechts2014_gasacc}. Once the planet core is large enough to perturb the gas in the disc, pebble accretion is halted and the planet can accrete a gaseous atmosphere by lowering its thermal energy. This mass is commonly referred to the pebble isolation mass \citep{bitsch2018_Miso}. Planets that have not yet reached pebble isolation mass after the dispersal of the disc have therefore not been able to accrete a significant atmosphere and could therefore plausibly constitute the peak at $\sim$1.5 $R_\oplus$. Further, the peak at the other end of the radius valley could be constituted by planets that have managed to reach pebble isolation mass and accrete a significant atmosphere. 

As pebble accretion has been shown to contribute to the formation of both gas giants \citep{bitsch2015_giants} and terrestrial planets \citep{johansen2021_pebSS,onyett2023_pebSS}, we will only focus on the mechanisms of pebble accretion during this work. We nevertheless note that a similar reasoning as above could be made when assuming instead that planets in the inner regions of the protoplanetary disc would grow by planetesimal accretion \citep{batygin2023_pltsacc}.

In this work, we aim to test whether we can create a primordial radius valley by simulating core formation by pebble accretion, including the accretion of gas after the cores reach pebble isolation mass. We then allow the planets that have accreted an atmosphere to contract without invoking mass loss and compare our synthesised planets to observations. In Sect. \ref{sec:core_formation} we briefly describe our core formation model while in Sects. \ref{sec:gas_acc} and \ref{sec:contract} we go over our gas accretion and contraction models. Finally we show our results in Sect. \ref{sec:pop_results}. We discuss and present our conclusion in Sect. \ref{sec:disc}. In Appendix \ref{app:v_frag}, we test the effects of using different fragmentation velocities for our pebbles. In Appendix \ref{app:disc_prop}, we show some properties of the protoplanetary discs in our model. In appendices \ref{app:lowalpha} and \ref{app:gap_disc}, we investigate using lower values for the viscous $\alpha$-parameter as well as including the emergence of a gap in the protoplanetary disc respectively. In Appendix \ref{app:E_c}, we derive the energy of the core of a planet and discuss its relevance in our model. Finally, in appendices \ref{app:photoevap}, \ref{app:steam}, and \ref{app:opacity_contraction}, we investigate the effects of photoevaporation, steam atmospheres, as well as different opacity values respectively. 
\section{Core formation}
\label{sec:core_formation}
In order to grow the planet cores, we use the pebble accretion model described in \citet{nielsen2023}. For details, we refer the readers there and only briefly go over the main points here. 

We calculate the growth of pebbles following \citet{drazkowska2021_pebbleflux} by assuming that the dust is initially micrometer-sized and grows by collisions. We limit the pebble sizes by taking into account radial drift, turbulence-driven fragmentation, and fragmentation from differential drift. Typical fragmentation velocities are usually set to between 1 m/s and 10 m/s \citep[see e.g.][]{gundlach_blum2015_vfrag,musiolik_wurm2019_sticking}. We nevertheless set the fragmentation velocity of pebbles to 2 m/s in order to reach Stokes numbers of approximately $\sim$0.01 in the outer regions of the discs\footnote{See figure 4 in \citet{nielsen2023} for the resulting Stokes numbers and pebble flux as a function of time at different locations in the protoplanetary disc.}. In Appendix \ref{app:v_frag}, we show the core mass distribution for our nominal model as well as when varying the fragmentation velocity for different pebble compositions. We find that for cores with masses $\lesssim$6 $M_\oplus$, the core mass distribution between the two models are similar, which means that our choice of fragmentation velocity will not have a major effect on the final result as we will mainly focus on planets that are less massive than this limit. We also take into account the sublimation of pebbles as they migrate across sublimation fronts (ice lines). Further, as we track the compositions of the cores as they grow by accreting pebbles, we are able to model their densities and therefore the sizes of the cores. 

Planets migrate through type-I migration, which we implement following the migration rate of \citet{kanagawa2018}. Once a planet has the pebble isolation mass $\Miso$, core accretion stops and gas accretion begins. The pebble isolation mass is given by \citet{bitsch2018_Miso} as 
\begin{equation}
\label{eq:Miso}
\begin{split}
    &M_{\rm iso} = 25\,M_\oplus\left(\frac{H/a}{0.05}\right)^3 \left[0.34\left(\frac{\log(\alpha_3)}{\log(\alpha_{\rm t})}\right)^4+0.66\right]\times
    \\
    &\left(1+\frac{\chi-2.5}{6}\right)\left(\frac{M_*}{M_\odot}\right),
\end{split}
\end{equation}
where $\alpha_3 = 10^{-3}$ is a constant, $H$ is the disc scale height, $\alpha_{\rm t}$ is a parameter setting the strength of the turbulence, $a$ is the distance to the star, and $\chi$ is the negative of the logarithmic pressure gradient. It can be defined from the temperature and surface density gradients and is set to $\chi\approx 2.786$. The the disc scale height is given by $H=c_{\rm s}/\Omega$, where $c_{\rm s}$ is the sound speed and $\Omega$ is the Keplerian frequency. We show both the aspect ratio, $H/a$, and pebble isolation mass as function of the distance to the host star for different times throughout the disc lifetime in Appendix \ref{app:disc_prop}. The sound speed is calculated from the ideal gas law
\begin{equation}
    c_{\rm s} = \sqrt{\rgas T},
\end{equation}
where $T$ is the temperature of the protoplanetary disc and $\mathcal{R} = k_{\rm b}/(m_{\rm H}\mu)$ is the specific gas constant, $k_{\rm b}$ is Boltzmann's constant, $m_{\rm H}$ is the hydrogen mass, and $\mu$ is the mean molecular weight of the gas, which we set to 2.35 for a mixture of H$_2$ and He. We calculate the temperature from a simple irradiated disc model from \citet{ida2016}
\begin{equation}\label{eq:t_irr}
    T_{\rm irr} = 150\left(\frac{L}{L_\odot}\right)^{2/7}\left(\frac{M_*}{M_\odot}\right)^{-1/7}\left(\frac{a}{\rm AU}\right)^{-3/7}.
\end{equation}
We assume that the accretion speed of the gas onto the star is set by the turbulent viscosity $\nu = \alpha c_{\rm s}H$, where $\alpha$ is the viscous $\alpha$-parameter. This results in a gas speed of $u_r = (3/2)\alpha c_{\rm s} (H/r)$. We distinguish the viscous $\alpha$ from $\alpha_{\rm t}$, setting $\alpha = 10^{-2}$ and $\alpha_{\rm t}=10^{-4}$. We specifically separate the parameter setting the strength of angular momentum transport, $\alpha$, and the parameter setting the strength of the turbulence and therefore the maximum pebble size in the fragmentation limit, $\alpha_{\rm t}$, since a single $\alpha$-parameter is not enough to model the possibility of laminar or weak-turbulence angular momentum transport of protoplanetary discs \citep{lesur2023_pp7discreview}. We choose to adopt a value of $\alpha = 0.01$, which have been needed in order to match the evolution of observed disc sizes \citep{najitabergin2018_alpha} as well as the decrease of stellar accretion rate with time \citep{hartmann1998}. We note however, that other observations have implied that $\alpha$ should be in the range of $\sim$10$^{-4}$-10$^{-3}$ \citep{trapman2020_alpha,rosotti2023_alphaobs}. For a full discussion on our choice of $\alpha$, we refer to Sect. 7.1 in \citet{nielsen2023}. In Appendix \ref{app:lowalpha}, we show our final planet population, similar to the top row in Fig. \ref{fig:atm_frac}, but for $\alpha=10^{-3}$.

We model the temporal decay of gas flux onto the star following \citet{hartmann1998}, keeping track of the amount of gas lost from the disc by accretion onto the star as well as from XUV photoevaporation using the mass-loss relation from \citet{owen2012_photoevap}. We model the XUV flux from the star as a function of stellar mass using the relation from \citet{bae2013_xray}. The inclusion of photoevaporation could cause an opening of a gap in the gas disc, which can influence the growth of planets. However, such an opening usually happens very late during the disc lifetime and therefore is expected to have a minimal effect on the planet population. We show this in Appendix \ref{app:gap_disc}. Similarly to \citet{nielsen2023}, we set the initial gas flux onto the star to be $\sim$$10^{-7}$ $M_\odot/{\rm yr}$, typical for solar mass stars. The lifetime of the disc is then equal to the time when all gas is lost. For a solar mass star, this becomes $\sim$2 Myr in our model. We set the initial gas mass of the disc to be $M_{\rm gas} = 0.1\,M_\odot$. We also vary the luminosity of the star during the lifetime of the disc according to the evolutionary tracks from \citet{baraffe2015_lum} for a solar mass star. 

\section{Gas accretion}
\label{sec:gas_acc}
\begin{figure*}
    \centering
    \includegraphics[width=\linewidth]{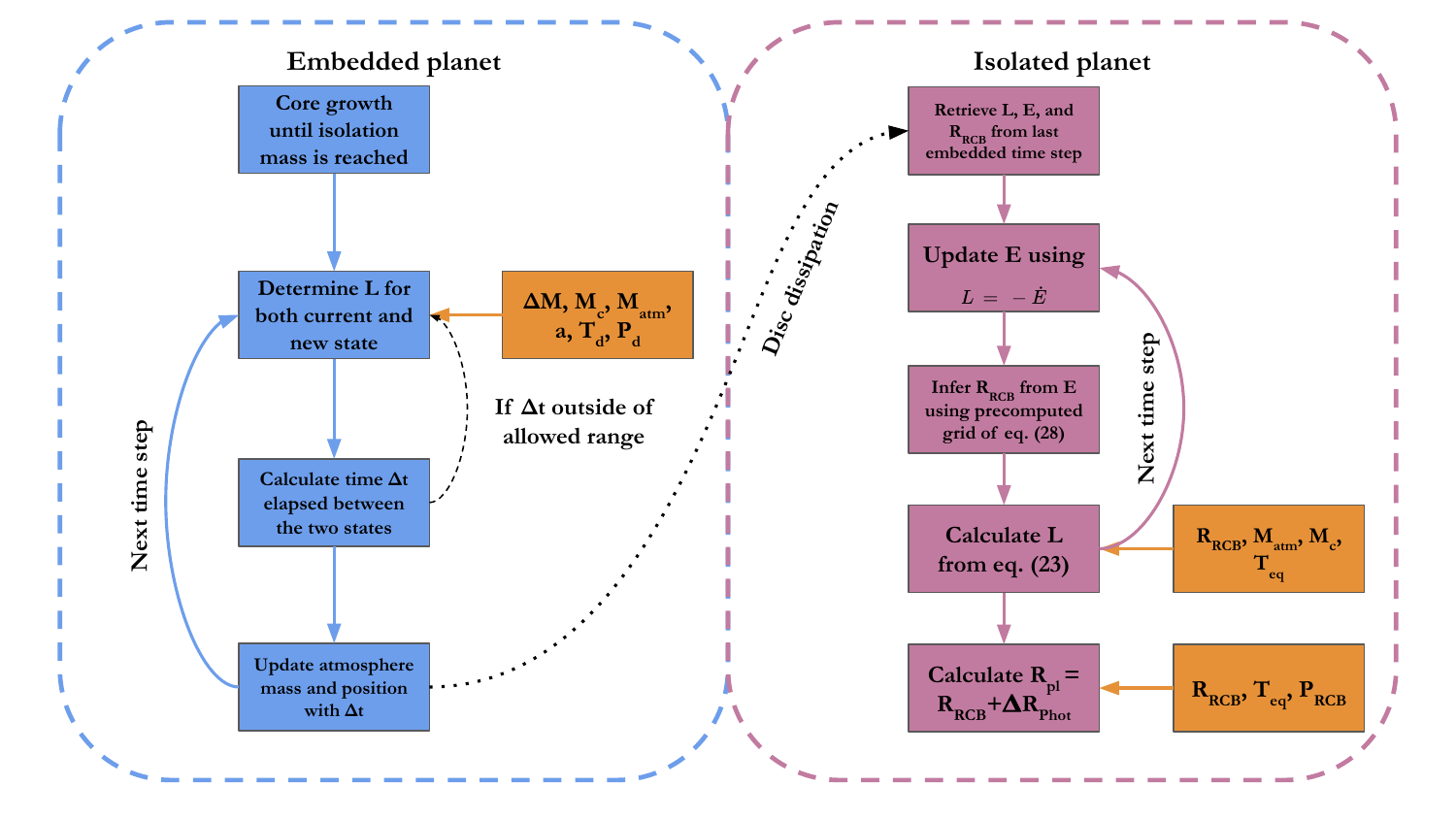}
    \caption{Flowchart of our procedure for calculating the accreted gas when the planet is embedded in the disc (left). We also show our method for calculating the rate of contraction after disc dissipation (right). The orange boxes show the input parameters going into each calculation. The main loops are shown as solid arrows in their respective colors. The dashed arrow indicates that we check if the calculated $\Delta t$ is within the allowed range and repeat the step with a new $\Delta M_{\rm tot}$ if it is not. The dotted arrow shows the transition from the embedded case to the isolated case after disc dissipation.}
    \label{fig:flowchart}
\end{figure*}

Once the pebble isolation mass is reached and pebble accretion is halted, the planet is able to cool down and accrete gas. In this section, we will describe our method for determining the structure of the atmosphere in order to find the luminosity of the planet, which in turn reduces the internal energy of the planet, cooling it down. As shown in previous work \citep{ikoma2000_gasacc,lee2014_method,lee2015_coolaccrete}, gas accretion is driven by atmospheres undergoing Kelvin-Helmholtz contraction from cooling, which allows more gas to become bound to the planet. We work under the assumption that the atmosphere is spherically symmetric and in hydrostatic equilibrium as well as in pressure balance with the midplane of the protoplanetary disc. We also assume that once the pebble isolation mass is reached and gas accretion has been initiated, the core is fixed in mass and radius i.e. it is not accreting any pebbles or planetesimals. In Fig. \ref{fig:flowchart} we show a flowchart of our procedure for calculating gas accretion when the planet is embedded as well as the contraction of the atmosphere after disc dissipation.
\subsection{Finding the luminosity}
\label{ssec:gas_acc_lum}
\begin{figure}
    \centering
    \includegraphics[width=\linewidth]{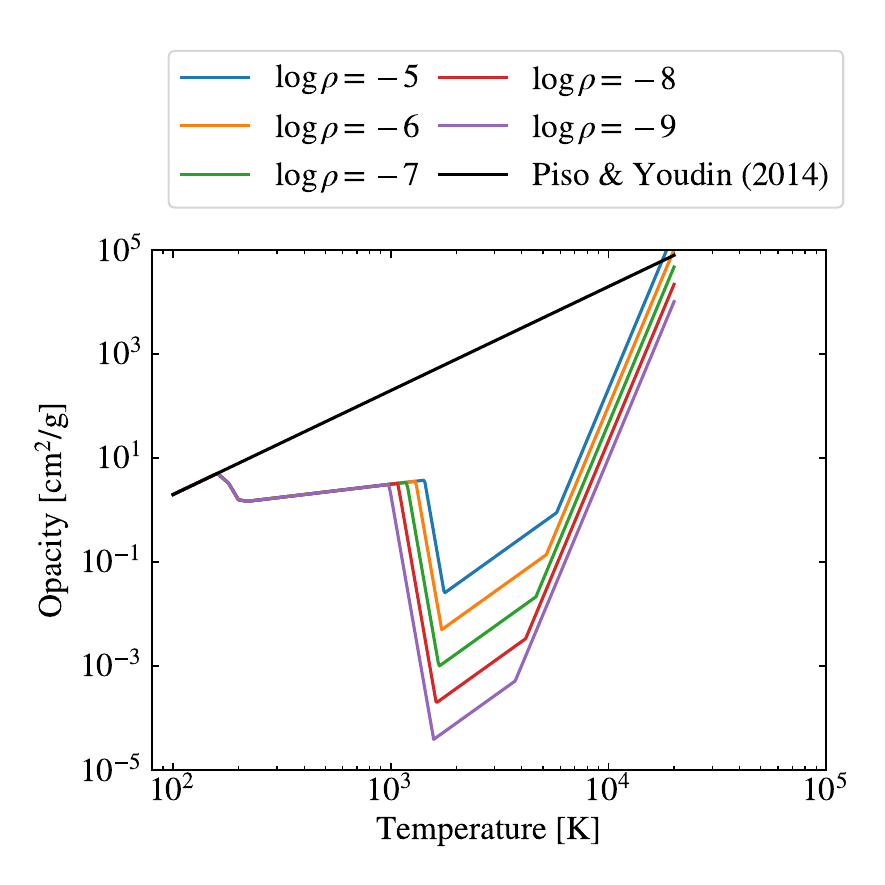}
    \caption{The opacities that we use in this work, shown for different gas densities in units of g/cm$^3$. We also show the opacity law used in \citet{pisoyoudin} using only ice grains as an opacity source. The main difference is caused by the evaporation of metal grains at $\sim$1000 K, which lowers the opacity significantly. The opacity decrease caused by the evaporation of ice grains happens at $\sim$200 K.}
    \label{fig:opacities}
\end{figure}

We find the luminosity from the two-layer\footnote{While the model of \citet{pisoyoudin} only includes two layers, we will show later on that our opacity model allows the formation of several radiative layers.} model of \cite{pisoyoudin}. The structure equations are
\begin{subequations}
\label{eq:structure}
    \begin{align}
        &\difffrac{m}{r} = 4\pi r^2\rho \label{eq:structure_m},\\
        &\difffrac{P}{r} = -\frac{Gm}{r^2}\rho \label{eq:structure_P}, \\
        &\difffrac{T}{r} = \nabla\frac{T}{P}\difffrac{P}{r},\label{eq:structure_T}
    \end{align}
\end{subequations}
where $G$ is the gravitational constant, $m$ is the enclosed mass at radius $r$, $P$ is the pressure, $T$ is the temperature, and $\rho$ is the density. We also assume that the gas can be seen as an ideal gas with equation of state
\begin{equation}
\label{eq:ideal_gas}
    P = \rho\mathcal{R}T.
\end{equation}
We assume that the atmosphere is divided in at least one inner convective and one outer radiative layer. The structure of the convective layer is independent of the luminosity $L$ and we assume that radiative layers have constant luminosity. The temperature gradient is set by $\nabla={\rm min}(\nablaad,\nabla_{\rm rad})$ where $\nabla_{\rm rad}$ is equal to
\begin{equation}\label{eq:nabla_rad}
    \nabla_{\rm rad} = \frac{3\kappa P}{64\pi G m \sigma T^4}L,
\end{equation}
where $\kappa$ is the opacity and $\sigma$ is Boltzmann's constant. Thus the boundary between the layers (the radiative-convective boundary, RCB) is set by $\nabla_{\rm rad} = \nablaad$. We assume that $\nablaad= 2/7$, which is true for an ideal diatomic gas with ratio of specific heats $\gamma=1/(1-\nablaad) = 1.4$. It is not possible to calculate the luminosity analytically without any further approximations. We therefore have to find it numerically. We initially guess a luminosity value and set the outer pressure and temperature boundaries to be equal to the disc conditions. The outer boundary condition for the mass is the total planet mass $\Mpl=\Mc+\Matm$. We then integrate the structure equations inwards until we reach the core radius $\Rc$. The luminosity is then iterated over until we find the correct luminosity which results in $m(\Rc) = \Mc$. The outer boundary $R_{\rm out}$ is set as the minimum between the Hill radius $R_{\rm H}$ and the Bondi radius $R_{\rm B}$ defined as
\begin{equation}
\label{eq:r_out}
    \begin{split}
        & R_{\rm H} = a\left(\frac{M_{\rm pl}}{3M_*}\right)^{1/3}, \\
        & R_{\rm B} = \frac{GM_{\rm pl}}{c_{\rm s}^2}.
    \end{split}
\end{equation}
We assume that the opacity in the radiative layer is set by dust grains and use
\begin{equation}\label{eq:opacity}
    \kappa = \kappa_i\rho^aT^b,
\end{equation}
as an opacity law. We take the parameters $\kappa_i$, $a$, and $b$ from \citet{bell_lin1994_opacities}, where we have assumed that the dominant sources of the opacity are ice grains, metal grains, and gas molecules. We show our opacity model in Fig. \ref{fig:opacities} where we also show the opacities used in \citet{pisoyoudin}, which only includes ice grains. It should be noted that in both models, $\trcb$ is at most $\sim$1000 K and below this, the atmosphere is fully convective and independent of the opacity. For clarity, we nevertheless show the opacities for the full temperature range of the atmosphere.

\begin{figure}
    \centering
    \includegraphics[width=\linewidth]{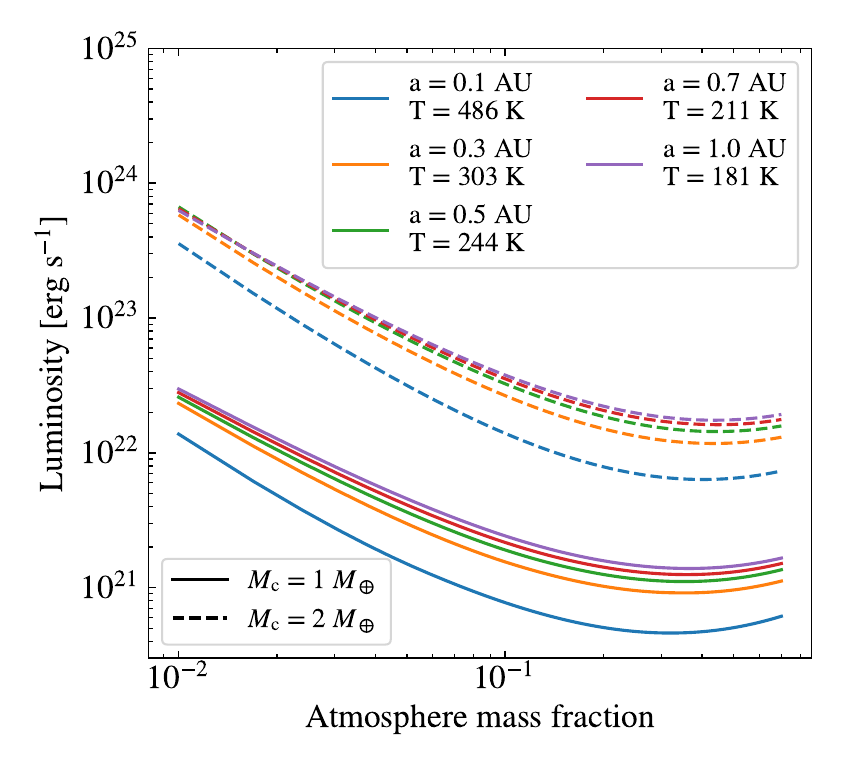}
    \caption{The embedded planet luminosity as a function of atmosphere mass fraction for different distances to the host star. We also show the temperature of the surrounding protoplanetary disc for all distances. The solid lines are the luminosities for a planet with a core mass of 1 $M_\oplus$ while the dashed lines are the luminosities using a planet with a core mass of 2 $M_\oplus$. For both core masses, the luminosity increases by a factor of $\sim$2.5 when placed at a distance of 1 AU compared to 0.1 AU. Further, when the core mass is doubled, the luminosity increases by a factor of $\sim$10 for similar distances to the host star.}
    \label{fig:lum_sma}
\end{figure}

We show the resulting luminosity for two different planet cores as a function of atmosphere mass fractions at several different distances to the host star in Fig. \ref{fig:lum_sma}. Clearly, increasing the core mass results in higher luminosities at all distances as a result of the increased pressures and temperatures in the atmosphere. Further, at larger distances to the host star, the luminosity increases. 

\subsection{Calculating the accreted mass}
\label{ssec:gas_acc_method}
The atmospheric energy (from assuming that the planet is in hydrostatic equilibrium with the disc) is the sum of the gravitational energy and the internal energy
\begin{equation}
    E = E_{\rm G} + U = -\int_{\Mc}^{M} \frac{Gm}{r}dm + \int_{\Mc}^{M} u dm,
\end{equation}
where the specific thermal energy $u$ can also be written as $\mathcal{R}T(\nablaad^{-1}-1)$. This integral has no analytical solution since the mass $m$ enclosed at radius $r$ is dependent on the radius and has no analytical form. Therefore, the accretion rate $\Dot{M}$ cannot be expressed analytically as a function of $L$ without neglecting the self-gravity of the atmosphere or the work done on the surface during accretion. \cite{pisoyoudin} solved this by considering the full energy equation at the RCB,
\begin{equation}\label{eq:glob_energy}
    L_{\rm M} = -\Dot{E} + e_{\rm M}\Dot{M}-P_{\rm M}\left.\frac{\partial V_{\rm M}}{\partial t}\right\rvert_{\rm M},
\end{equation}
where the index $M$ means that the quantity is evaluated at surface $R$, where the enclosed mass $m$ is equal to $M$. The specific energy $e_{\rm M}$ at this boundary is simply $u_{\rm M}-GM/R$. Here, we have chosen to neglect any heating from the accretion of solids such as the accretion of planetesimals after the planet has reached pebble isolation mass as well as any potential heating from radioactive decay. Including radioactive heating would slow down the cooling of the entire atmosphere. However, models including heating from radioactive decay have found that this has a negligible effect on the overall luminosity of the planet as the radiogenic energy release rate is typically orders of magnitude lower than the luminosity from radiation as described in this work \citep{lopez2012_massloss,owenwu2016_boiloff}. 

For a given atmosphere mass and a set of boundary conditions for $T$ and $P$, the structure of the atmosphere, including the total atmospheric energy $E$, is uniquely determined by the luminosity $L$. From eq. \eqref{eq:glob_energy}, we can also see that the luminosity in turn sets the accretion rate $\Dot{M}$. The elapsed time $\Delta t$ between two different mass states $i$ and $i+1$ can therefore be expressed as
\begin{equation}\label{eq:delta_t}
    \Delta t = \frac{-\Delta E +\avg{e}\Delta M - \avg{P}\Delta V_{\avg{M}}}{\avg{L}},
\end{equation}
where $\avg{}$ denotes the average between state $i$ and $i+1$ while $\Delta$ denotes the difference. This method makes it possible to calculate the time it takes to accrete a certain mass. At each time step, we therefore integrate the planet atmosphere with mass $M_{i}$, retrieving the luminosity by iteratively matching to the desired mass. We then choose a new atmosphere mass $M_{i+1} = M_{i}+\Delta M_{\rm tot}$ and integrate the atmosphere again, using the updated atmosphere mass. We note the difference between $\Delta M_{\rm tot}$ and $\Delta M$: $\Delta M$ is the mass difference at the height $R$ where we choose to evaluate eq. \eqref{eq:delta_t} while $\Delta M_{\rm tot}$ is the total mass difference of the atmospheres. The difference in energy and mass between state $i+1$ and state $i$ together with the luminosity at state $i$ allows us to calculate the time elapsed between the two states through equation \eqref{eq:delta_t}. As $\Delta t$ sets the time step of the integration, it is necessary to prevent too small or too large time steps. We therefore only allow $\Delta t$ to vary between 1 and 3000 years. Should $\Delta t$ be outside of this range, we retry the integration with a different $\Delta M_{\rm tot}$.

The accretion rate from contraction may become higher than the rate at which gas can be supplied to the planet from the disc. In that case, we use the gas accretion scheme of \citet{ida2018} who found that the gas enters the Hill sphere at a rate of
\begin{equation}
\begin{split}
    &\left.\frac{{\rm d}M}{{\rm d}t}\right\rvert_{\rm Hill} = 1.5\times 10^{-3}\,M_\oplus\,{\rm yr^{-1}}\left(\frac{H/a}{0.05}\right)^4\times
    \\
    &\left(\frac{M}{10M_\oplus}\right)^{4/3}\left(\frac{\alpha}{0.01}\right)^{-1}\left(\frac{\Dot{M}_{\rm g}}{10^{-8}M_\odot{\rm yr^{-1}}}\right)\frac{1}{1+(M/M_{\rm gap})^2},
\end{split}
\end{equation}
where $M_{\rm gap}$ is the mass required to carve a gap in the protoplanetary disc\footnote{This occurs at $\Mpl$$\sim$2.3 $M_{\rm iso}$ \citep{johansen2019}.}, and $\Dot{M}_{\rm g}$ is the gas flux in the protoplanetary disc. Should the contraction rate be faster than the rate at which gas enters the Hill sphere, the Hill accretion rate will set the gas accretion rate. Finally, both of these processes are limited by the gas flux onto the star $\Dot{M}_{\rm g}$. We therefore set the final gas accretion rate to be 
\begin{equation}
    \difffrac{M_{\rm atm}}{t}={\rm min}\left(\frac{\Delta M_{\rm tot}}{\Delta t},\left.\difffrac{M}{t}\right\rvert_{\rm Hill},\Dot{M}_{\rm g}\right).
\end{equation}
Given that the opacity decreases during sublimation of ice grains and dust (see Figure \ref{fig:opacities}), it is possible for radiative windows to exist within the atmosphere. We therefore evaluate eq. \eqref{eq:delta_t} at the outermost RCB as the luminosity in the innermost RCB can underestimate the total luminosity should the radiative windows be too large \citep{piso2015}.

\begin{figure}
    \centering
    \includegraphics[width=\linewidth]{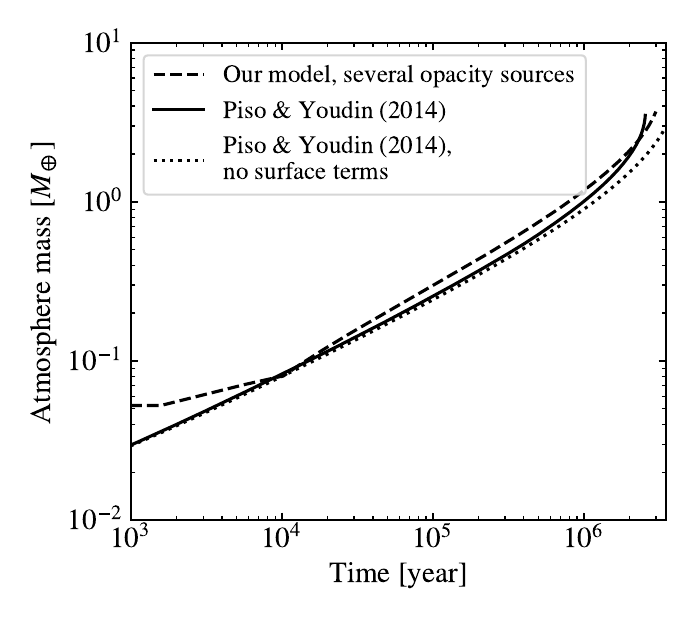}
    \caption{The accreted gas mass as a function of time for a 5 $M_\oplus$ core orbiting at 60 AU using the same disc model as \citet{pisoyoudin}. We also show a reproduction of their model. When taking into account a more complex opacity model based on \citet{bell_lin1994_opacities}, resulting in lower opacity due to ice and silicate dust sublimation, gas accretion is slightly more efficient due to the increased luminosities. We also show the case when the surface terms in eq. \eqref{eq:delta_t} are ignored which results in slightly slower gas accretion.}
    \label{fig:gas_acc_compare}
\end{figure}

Figure \ref{fig:gas_acc_compare} shows our gas accretion method in comparison to the method by \citet{pisoyoudin} for a 5 $M_\oplus$ core at 60 AU using the same protoplanetary disc boundary conditions as in their work but with detailed opacities that allow metal grains and gas molecules to affect the opacity as well. Using the more complex opacity law results in slightly faster gas accretion due to the lower opacities and therefore higher luminosities. 
\subsection{Calculating the atmosphere mass during pebble accretion}
\label{ssec:gas_acc_init}
\begin{figure}
    \centering
    \includegraphics[width=\linewidth]{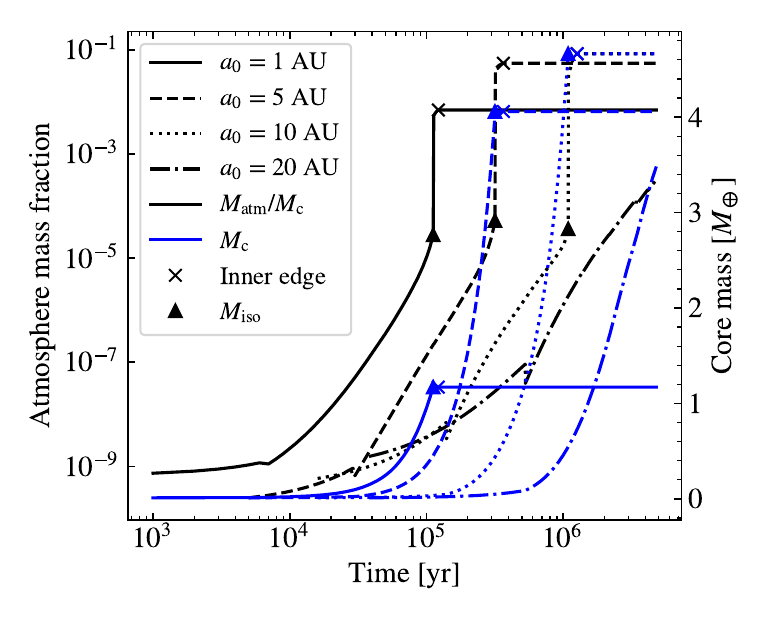}
    \caption{Atmosphere mass fraction and core mass for four planets as a function of time. The atmosphere masses remains low during pebble accretion due to the high accretion luminosity. Once pebble accretion is halted after pebble isolation mass is reached (triangles), the gas rapidly contracts due to the high initial luminosity. With the exception of the planet injected at 20 AU (dash-dotted line), all planets reach the inner edge (crosses), where gas accretion is halted, before a massive atmosphere has been accreted.}
    \label{fig:atm_frac_time}
\end{figure}

Any planet embedded in a protoplanetary disc will have some gas bound to it. However, infalling pebbles will deposit energy into the atmosphere, heating it up and preventing significant contraction of the gas. This causes the accretion of gas to be slowed down significantly \citep{lambrechts2014_gasacc,lambrechts2017_gasacc}. During pebble accretion, the bound atmosphere mass of a protoplanet is uniquely determined by the disc conditions and the accretion luminosity of the planet. The accretion luminosity is given by
\begin{equation}
\label{eq:L_acc}
    L_{\rm acc} = \frac{G\Mc\Dot{M}}{\Rc},
\end{equation}
where $\Dot{M}$ is the accretion rate of pebbles, which we calculate according to equation (2) in \citet{nielsen2023}. We then use the analytical model of \citet{pisoyoudin} to determine atmosphere mass of the protoplanets where
\begin{equation}
    \Matm = \frac{\prcb\Mc}{\xi P_{\rm M}},
\end{equation}
and
\begin{subequations}
\label{eq:M_atm_const}
    \begin{align}
        &\xi = \sqrt{\ln(\prcb/(\theta P_{\rm d}))},
        \\
        &P_{\rm M} = \frac{4T_{\rm d}^4\chi^{7/2}\rgas^4}{5\pi^2G^3\Mc^2\nablaad^{5/2}},
    \end{align}
\end{subequations}
where $\chi$ and $\theta$ are constants arising from the fact that the outermost radiative layer is not isothermal; we set these constants to 1.53 and 0.557 respectively, following \citet{pisoyoudin}. We approximate the pressure at the radiative-convective boundary (defined further down), $\prcb$, as 
\begin{equation}
\label{eq:Prcb}
    \frac{\prcb}{P_{\rm d}} = \frac{\nablaad/\nabla_{\rm rad}}{1-\nablaad/\nabla_\infty}+1,
\end{equation}
where $\nabla_\infty = 1/(4-\beta)$ and $\beta$ is the temperature power-law index of the opacity. Given the low temperatures in the disc and upper regions of bound atmospheres, we assume that the opacity in these cases is dominated by ices and set $\beta=2$. We use the pebble accretion luminosity, $L_{\rm acc}$, in order to calculate the radiative temperature gradient, $\nabla_{\rm rad}$, in equation \eqref{eq:Prcb}. We note that this analytical model is only valid for very low atmosphere masses and is therefore only used for planets that have not yet reached pebble isolation mass. Once a planet reaches pebble isolation mass, we turn to our gas accretion model described in Sect. \ref{ssec:gas_acc_method}. We show the atmosphere mass fraction as a function of time for four planets with different starting locations in a disc around a star with solar metallicity and solar mass in Fig. \ref{fig:atm_frac_time}. During pebble accretion, the atmosphere mass remains low as a result of contraction being slow. However, as the mass of the core grows, so does the bound atmosphere mass. Once pebble accretion is halted, the accretion of gas is initially very efficient as the luminosity of the planet is high, which results in rapid contraction. Due to this rapid accretion in the early stages after pebble isolation mass is reached, the planet population is naturally separated into planets with very low atmosphere mass fractions and planets with atmosphere mass fractions of a few percent. All planets except the one injected at 20 AU migrate to the inner edge, where gas accretion is halted.

\section{After disc dissipation}
\label{sec:contract}
After the dissipation of the disc, the planet surroundings are laid bare and the previously used disc boundary conditions are no longer relevant. These pressure and temperature boundary conditions are instead replaced by mass and energy conservation, which are integral quantities not easily connected to the luminosity without distinct boundary conditions. We therefore take an analytical approach in order to find the luminosity and thus the contraction rate of the planet after the dissipation of the protoplanetary disc. The main difference between this analytical model and the numerical model in Sect. \ref{sec:gas_acc} is that we replace the mass $m(r)$ enclosed at radius $r$ with the total planet mass $\Mpl$ (including the atmosphere) in order to be able to solve the integrals over the density and temperature profiles of the atmosphere. 

Previous works have mostly focused on the effect of atmospheric mass-loss following the loss of pressure support from the surrounding protoplanetary disc and exposure to stellar radiation as a means of shaping the radius valley \citep{owen_wu2017_PE,ginzburg2018_cpm}. In this work, we aim to explore the formation of a primordial radius valley without including any form of mass-loss, such as spontaneous mass loss known as ''boil-off'', which occurs due to the loss of pressure support from the protoplanetary disc \citep{owenwu2016_boiloff}. We briefly discuss this omission of the boil-off effect in Sect. \ref{ssec:disc_photoevap}. By this omission, we implicitly assume that the contraction rate of the atmosphere is faster than the expansion caused by heating from either the stellar XUV flux or thermal energy of the planet. The goal of this section is therefore to find an analytical expression for the total atmospheric energy of the planet as a function of the height of the RCB, which in turn can be used to estimate the radius of the planet. At the final time step before disc dissipation, we find the height of the RCB of the planet which we use to analytically find the luminosity and the energy at the first time step after disc dissipation. Given that the planet is no longer embedded in a disc, the surface terms in eq. \eqref{eq:glob_energy} are equal to zero as there is no work done on the atmosphere anymore and the atmosphere does not accrete any more mass. Instead, the only change in energy comes from radiative heat loss and subsequent contraction. The energy of the planet evolves through $L = -\Dot{E}$ and thus we can infer the height of the RCB over time. This method is similar in nature to previous work by \citet{owen_wu2017_PE,leeconnors2021_primordial}. 
\subsection{Finding the luminosity}
\label{ssec:contract_lum}
The density profile of the innermost convective region can be written as
\begin{equation}
\label{eq:rho_rcb_anal}
    \rho = \rhorcb\left(1+\frac{\rbprime}{r}-\frac{\rbprime}{\rrcb}\right)^{1/(\gamma-1)}.
\end{equation}
Here we have defined\footnote{We define the effective Bondi radius $\rbprime$ based on the total planet mass $\Mpl$, including the total atmosphere. This results in a slight overestimation of the effects of self-gravity and thus of the luminosity and total energy of the planet. We choose to use $\Mpl$ instead of $\Mc$ however, in order to not fully neglect self gravity.} the effective Bondi radius $\rbprime$. 
\begin{equation}
\label{eq:R_B_prime}
    \rbprime = \nablaad\frac{G\Mpl}{\rgas\trcb}.
\end{equation}
We assume that the temperature at the RCB, $\trcb$, is related to the equilibrium temperature of the planet $T_{\rm eq}$. The latter is defined as
\begin{equation}
\label{eq:T_eq}
    T_{\rm eq} = \left(\frac{L_*}{4\pi\sigma a^2}\right)^{1/4},
\end{equation}
where $a$ is the semi-major axis of the planet. We set $\trcb = \chi T_{\rm eq}$. We find the atmosphere mass by integrating the density profile from the core up until the RCB, making use of the assumption that the radiative layer holds a negligible amount of the atmosphere mass. This yields
\begin{equation}
\label{eq:Matm}
\begin{split}
    \Matm = & \int_{\Rc}^{\rrcb}4\pi r^2 \rho dr = 4\pi \rhorcb I_2,
\end{split}
\end{equation}
where we define the integral
\begin{equation}
    I_n = \int_{\Rc}^{\rrcb} r^n \left(1+\frac{\rbprime}{r}-\frac{\rbprime}{\rrcb}\right)^{1/(\gamma-1)} dr.
\end{equation}
The analytical solution to $I_n$ only has real solutions when $\rrcb < \rbprime$, which is not a physical limitation. We therefore choose to solve this integral numerically. At the RCB, we know that $\nablaad=\nabla_{\rm rad}$, meaning we can write the luminosity from equation \eqref{eq:nabla_rad} as
\begin{equation}
\label{eq:luminosity_anal}
\begin{split}
    L = &\, \frac{64\pi\sigma_{\rm sb}\trcb^4\rbprime}{3\kappa_{\rm RCB}\rhorcb} = \frac{256\pi^2\sigma_{\rm sb}\trcb^4\rbprime I_2}{3\kappa_{\rm RCB}\Matm},
\end{split}
\end{equation}
where we have inserted $\rhorcb$ from equation \eqref{eq:Matm}. We now have an expression for the luminosity as a function of the atmosphere mass, given a specific height and temperature of the RCB.
\subsection{Connecting the energy to the radius}
\label{ssec:contract_energy}
\begin{figure}
    \centering
    \includegraphics[width=\linewidth]{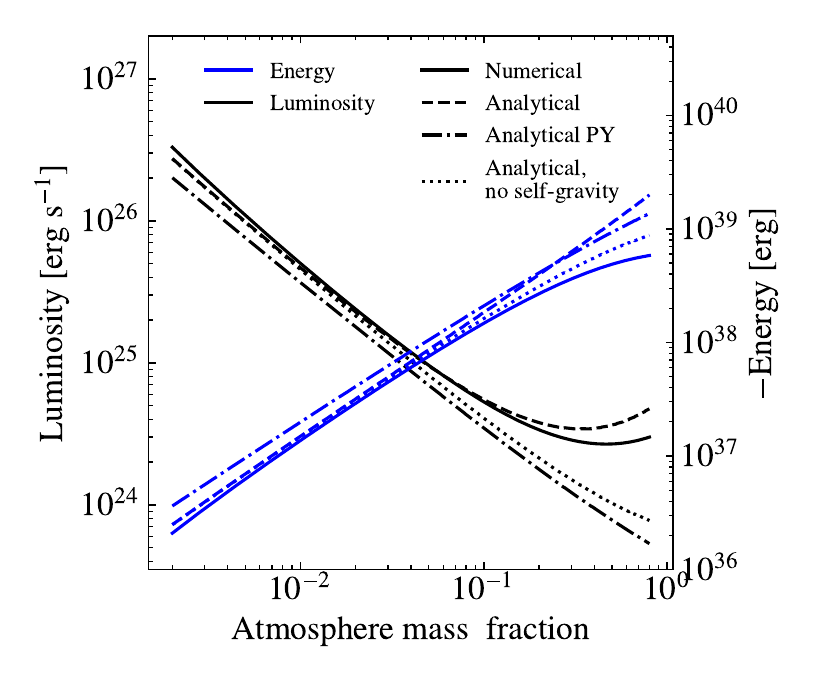}
    \caption{Comparison of luminosity (black) and total atmospheric energy (blue), between the full numerical solution (solid) presented in Sect. \ref{ssec:gas_acc_lum} and our analytical model (dashed) for the same setup as in Fig. \ref{fig:gas_acc_compare}. We also show the analytical model presented by \citet{pisoyoudin} i.e. their eqs. (30c) and (31) (dash-dotted) and our analytical model with no self-gravity (dotted). While the atmospheric energies are similar, by including the entire planet mass, we overestimate the effects of self-gravity, especially at higher masses. This leads to an increased luminosity at higher atmosphere masses. The analytical model in \citet{pisoyoudin}, in turn, neglects self-gravity completely and thus underestimates the luminosity. Our analytical model with self-gravity turned off results in the luminosity being marginally higher than the analytical model in \citet{pisoyoudin} but with a similar shape.}
    \label{fig:analytical_comp}
\end{figure}

The specific energy $e$ of a parcel of gas in the atmosphere can be written as
\begin{equation}
\label{eq:specific_energy}
    e = u+e_g = \rgas T(\nablaad^{-1}-1)-\frac{G\Mpl}{r},
\end{equation}
where we again note the replacement of the enclosed mass $m$ at radius $r$ with the total planet mass $\Mpl$. The temperature profile of the convective region, in turn, can be written as
\begin{equation}
\label{eq:T_prof}
    T = \trcb\left(1+\frac{\rbprime}{r}-\frac{\rbprime}{\rrcb}\right).
\end{equation}
This means that we can retrieve the total energy of the atmosphere by integrating the specific energy over the mass of the atmosphere,
\label{eq:E_analytical}
\begin{equation}
\begin{split}
    E = & \int_{\Rc}^{\rrcb} 4\pi r^2\left(\rgas T(\nablaad^{-1}-1)-\frac{G\Mpl}{r}\right)\rho dr = \\
    & \left(\rgas\trcb\frac{1-\nablaad}{\nablaad}-(1-\nablaad)\frac{G\Mpl}{\rrcb}\right)\int_{\Rc}^{\rrcb} 4\pi r^2\rho dr \\
    & - \nablaad G \Mpl\int_{\Rc}^{\rrcb}4\pi r \rho dr = \\
    & \Matm\rgas\trcb\frac{1-\nablaad}{\nablaad}\left(1-\frac{\rbprime}{\rrcb}\right)-4\pi\rbprime\rgas\trcb \rhorcb I_1.
\end{split}
\end{equation}
Inserting $\rhorcb$ from equation \eqref{eq:Matm}, we then get an expression for the energy
\begin{equation}
\label{eq:E_atm_full}
\begin{split}
    E = \Matm\rgas\trcb\frac{1-\nablaad}{\nablaad}\left(1-\frac{\rbprime}{\rrcb}\right) - \rbprime\rgas\trcb\Matm\frac{I_1}{I_2}.
\end{split}
\end{equation}
In Fig. \ref{fig:analytical_comp}, we compare this analytical model to the numerical model we present in Sect. \ref{ssec:gas_acc_lum} for an embedded planet with the same conditions as in Fig. \ref{fig:gas_acc_compare}. We use $\rrcb$ from the numerical integration of the atmosphere structure equations \eqref{eq:structure_m}-\eqref{eq:structure_T} as input in the analytical model. We also show the analytical model from \citet{pisoyoudin}, which neglects self-gravity and assumes $\Rc \ll \rrcb \ll \rbprime$. Further, we also show our model with neglected self-gravity, i.e. with the effective Bondi radius $\rbprime$ defined using only the core mass $\Mc$ instead of the mass of the entire planet $\Mpl$. While our analytical model overestimates the luminosity as expected, the luminosity has a similar shape as the numerical solution with a nearly flat relation at higher atmosphere masses. This can be contrasted to the two models neglecting self-gravity where the luminosity decreases with atmosphere mass. Our analytical model with self-gravity turned off results in marginally higher luminosities compared to the model of \citet{pisoyoudin}. For most atmosphere masses, the absolute value of the energy agrees well with the numerical solution for all analytical models. The numerical model is under all circumstances unpractical to use for the non-embedded atmosphere case, because the disc boundary conditions have been replaced by two integral conditions for total mass and total energy. 

Although we have defined both the atmosphere mass and the energy as a function of the height of the RCB, the height of the RCB can not be analytically determined. We can however rewrite eq. \eqref{eq:E_atm_full} by isolating all terms containing $\rrcb$ such that
\begin{equation}
\label{eq:grid_eq}
\begin{split}
    \frac{E}{\Matm\rgas\trcb} = \frac{1-\nablaad}{\nablaad}\left(1-\frac{\rbprime}{\rrcb}\right) -\rbprime\frac{I_2}{I_1}.
\end{split}
\end{equation}
We recall that $\trcb$ is a function of the equilibrium temperature only (equation \eqref{eq:T_eq}) and that $\rbprime$ is a function of $\trcb$ and total planet mass ($\Mc+\Matm$). The first term on the right hand side in equation \eqref{eq:grid_eq} is negligible when $\rrcb\sim\rbprime$ since the temperature at the bottom of the atmosphere is disconnected from the height of the RCB. This is true soon after disc dispersal. However, when contraction occurs and $\rrcb$ starts to become smaller than $\rbprime$, this term will become relevant and slow down contraction, meaning that it is necessary to include it in order to accurately describe the contraction of the atmosphere. We choose to neglect the thermal energy of the core when estimating the total energy budget of the planet. The temperatures at the bottom of the atmosphere are high enough for the core to be completely supercritical, convective, and nearly isothermal due to the low compressibility of the magma, similar to previous work \citep[e.g.][]{ginzburg2016_cpm}. This, together with the fact that the temperature at the bottom of the atmosphere varies very little during contraction causes the cooling of the core to have little to no effect on the cooling on the atmosphere. We show the derivation of the total energy of the core in Appendix \ref{app:E_c}. The energy of the core scales as $\left(\frac{1}{\Rc}-\frac{1}{\rrcb}\right)$, which means that for $\rrcb \gg \Rc$, this term can be approximated as constant with respect to $\rrcb$ and will not play a role in the contraction of the atmosphere. The core energy depends on the height of the RCB due to the fact that the temperature on the surface of the core is equal to the temperature at the bottom of the atmosphere. However, for large enough $\rrcb$, the temperature at the bottom of the atmosphere flattens out and becomes approximately constant.

We create a grid with different values of $\rbprime$, $\Rc$, and $\rrcb$ and retrieve $\rrcb$ from equation \eqref{eq:grid_eq} for given values of $E$, $\Rc$, and $\rbprime$. At each time step we then numerically evolve the energy from $L = -\Dot{E}$. As $\rbprime$ and $\Rc$ of all planets are known quantities, we can thus match the calculated atmospheric energy to some $\rrcb$ and track the evolution of the height of the RCB. Finally, the observed transit radius is not necessarily the radiative convective boundary or the photospheric radius (where optical depth, $\tau=2/3$) but rather where the optical depth for emitted stellar photons tangential to the atmosphere is unity, typically a few atmospheric scale heights above the RCB. The pressure at this radius is typically found to be $\sim$$10-30$ mbar \citep{burrows2004_transitradius,lopezfortney2014_inflate}. We note that the pressure at the transit radius might vary for different planets, resulting in larger or smaller than planets than estimated using a transit pressure of 20 mbar. For simplicity, we nevertheless adopt this value for the transit pressure and set the planet radius to be equal to $\Rpl = \rrcb+\Delta R_{\rm tr}$, where $\Delta R_{\rm tr}$ is equal to
\begin{equation}
\label{eq:R_tr}
    \Delta R_{\rm tr} = H_{\rm pl}\log\left(\frac{\prcb}{20\,{\rm mbar}}\right) = \frac{\rgas \trcb}{g}\log\left(\frac{\prcb}{20\,{\rm mbar}}\right),
\end{equation}
where $H_{\rm pl}$ is the scale height of the atmosphere and $g=G\Mpl/\rrcb^2$ is the gravity at the RCB.
\section{Planet population results}
\label{sec:pop_results}
We simulate core formation and atmosphere accretion in a protoplanetary disc around a solar mass host star according to the procedure described in Sects. \ref{sec:core_formation} and \ref{sec:gas_acc}. We assume that the host star is of solar metallicity, by implementing elemental abundances from \citet{magg2022_solarcomp}, resulting in a host star metal mass fraction $Z = 0.014$. The initial protoplanet mass is set to $0.01\, M_\oplus$. We sample the injection time of protoplanets uniformly between 1 kyr and 1 Myr and injection location from a log-uniform distribution between 0.5 and 20 AU. We tested several different outer and inner limits for both injection location and injection time but found little to no change in the final atmosphere mass fraction distribution and core mass distribution. After disc dispersal, we let the planets radiatively cool and contract over a timespan of 3 Gyr. We perform three sets of simulations. In the first, we let the planets migrate until they reach the inner edge of the disc, which we fix at 0.1 AU. We then stop any form of accretion at the inner edge. We perform a second simulation where planets still migrate to the inner edge but where we allow pebble accretion, as well as contraction and gas accretion after the pebble isolation mass is reached. Finally, we perform a simulation where we set 7 different inner edges spaced so the orbital period of a planet at each inner edge is in a 3:2 resonance with their inner neighbour. This approach is motivated by previous work, finding that inner planetary systems are expected to get locked into resonances by migration \citep{izidoro2017_resonance,kajtazi2023_resonance}. The innermost edge remains at 0.1 AU. We then assign each planet one of these inner edges randomly as stopping point. The planets then grow and migrate inwards until they reach their assigned inner edges, after which they continue to grow through pebble and gas accretion. We also do not consider gas accretion for planets with masses lower than 1 $M_\oplus$ in order to speed up computations. The isothermal masses for these planets are low enough such that they would not accrete a significant atmosphere, which means that we can neglect gas accretion for these planets \citep{lee2022_gapwithoutmassloss}.
\subsection{Atmosphere masses}
\label{ssec:pop_mass}
\begin{figure*}
    \centering
    \includegraphics[width=\linewidth]{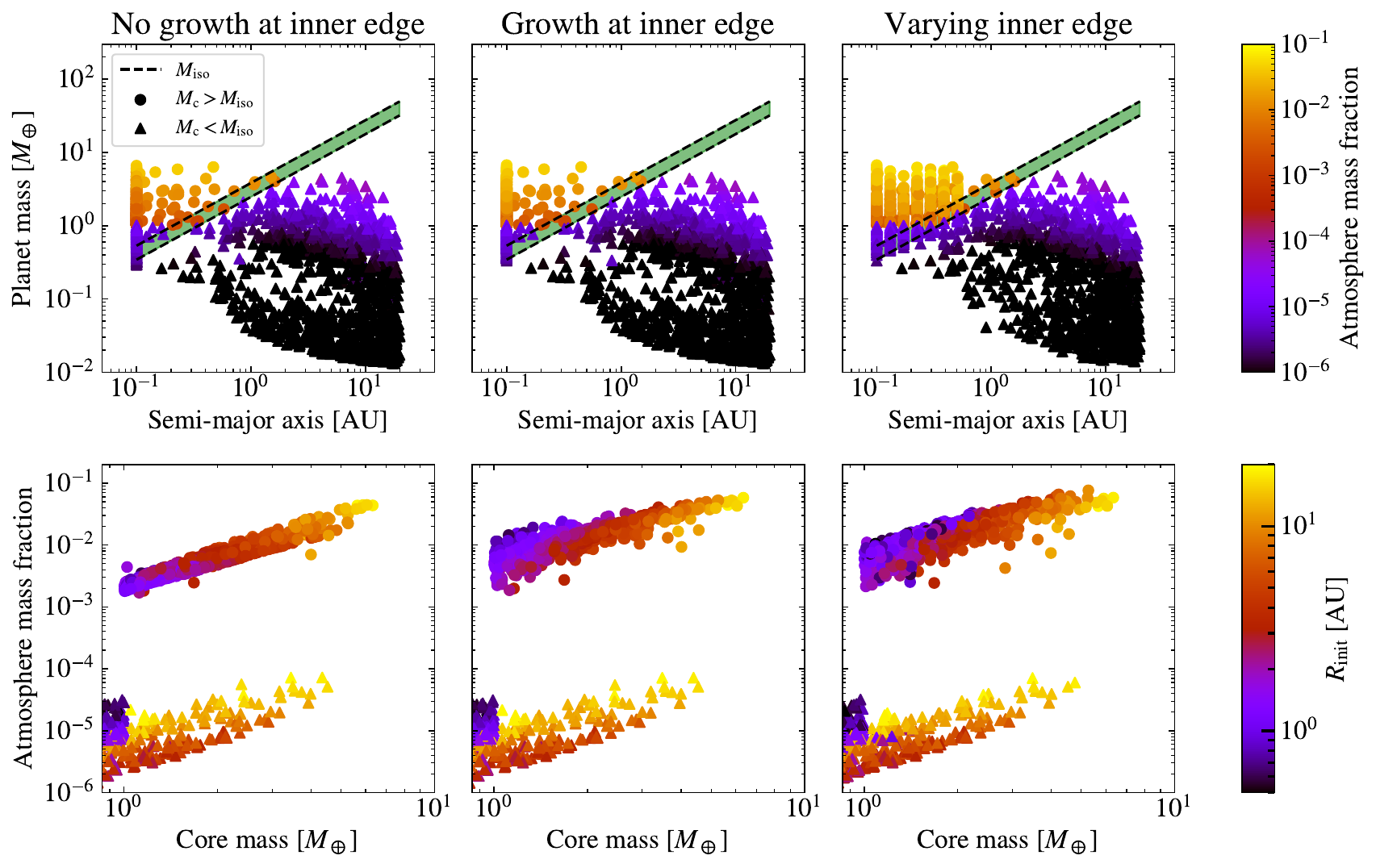}
    \caption{Top row: Final semi-major axes and planet masses for the planet population after disc dissipation. The color denotes the final atmosphere mass fraction. The dashed lines show the pebble isolation mass at the start and end of the disc lifetime; the variation is due to the temporal change of the stellar luminosity. Bottom row: Atmosphere mass fraction as a function of core masses. The colors of the points show the injection location in the disc. In the left column we show the the population when gas accretion at the inner edge is halted. In the middle column we show the population when gas accretion continues at the inner edge while in the right column we show the planet population when we vary the inner edge. In general, atmosphere mass fractions for massive cores are similar in all three models. If gas accretion is halted at the inner edge, atmosphere mass fractions are low (up to $\sim$0.1-5\%) due to the efficient migration happening in the disc. When gas accretion continues at the inner edge, the atmosphere mass fractions of the smallest cores can reach up to 1\%. Finally, when varying the inner edge, planets that are accreting gas further away from the star are able to accrete more gas due to more efficient contraction, reaching atmosphere mass fractions of almost $\sim$10\%.}
    \label{fig:atm_frac}
\end{figure*}

We present the resulting planet population in Fig. \ref{fig:atm_frac}. In the top row we show the resulting semi-major axes and planet masses for all planets that were injected. The colour on each point shows the atmosphere mass fraction of each planet. We also show the pebble isolation mass during the disc lifetime in between the two dashed lines\footnote{As the pebble isolation mass depends on the scale height of the disc, it decreases over time as the stellar luminosity decreases.}. In the bottom row we show the atmosphere mass fractions as a function of their core mass. Here, the colour indicates the injection location of the planet while the grey points show the initial atmosphere mass fractions, which we calculate according to the description in Sect. \ref{ssec:gas_acc_init}. The left column shows the planet population when gas accretion is halted at the inner edge while the middle column shows the results when gas accretion is allowed to continue at the inner edge. The right column shows the population when we vary the inner edge. As most planets do not reach the pebble isolation mass, they cannot accrete a significant atmosphere and therefore remain small. Indeed, planets with core masses below the pebble isolation mass have atmosphere mass fractions $\lesssim$10$^{-5}$. For such low atmosphere mass fractions, we can approximate $\Rpl\sim\Rc$ \citep{lee2022_gapwithoutmassloss}.

In all cases where the core mass is higher than the pebble isolation mass, atmosphere mass fractions remain below $\sim$10\%, in line with estimations for non-giant planets observed by the Kepler satellite \citep{lopezfortney2014_inflate,wolfganglopez2015_rocky,rogers_owen2021_initpop} as well as previous planet population synthesis models \citep{mordasini2020_popsynth}. In the first model, where gas accretion is halted at the inner edge, the atmosphere mass fractions vary between $\sim$0.2\% to $\sim$5-6\% for core masses between 1 $M_\oplus$ and 6 $M_\oplus$. When allowing gas accretion to occur at the inner edge, efficient migration is not hindering growth as much and lower mass planets (1-2 $M_\oplus$) reaching the inner edge early are able to accrete slightly more gas, reaching, for these low-mass planets, atmosphere mass fractions as high as $\sim$3\%. As the planet luminosity decreases closer to the star, the accretion of gas is relatively inefficient in this region. The atmosphere mass fractions for high core masses is therefore similar in the two models where the inner edge is located at 0.1 AU. 

Planets that have their migration halted further away from the star accrete gas more efficiently as they have higher luminosities and their atmospheres contract faster. When varying the inner edge, a few massive cores are able to reach atmosphere mass fractions up to 10\%. Further, in this model, planets with masses $\sim$2-3 $M_\oplus$ are able reach atmosphere mass fractions of $\sim$4-5\%. Only cores with masses $\sim$5-6 $M_\oplus$ were able to reach such high atmosphere mass fractions in the case where accretion is halted at the inner edge. In the second case, with the inner edge at 0.1 AU and gas accretion continues, only cores with masses $\gtrsim$4 $M_\oplus$ could reach such high atmosphere mass fractions. Clearly, migration plays an important role in shaping the final masses of the planets as it puts constraints on how much gas a planet has time to accrete. Even when accretion is not halted at the inner edge, the position of the inner edge will affect how efficient atmosphere contraction is and therefore the amount of gas a planet can accrete.

Further, almost all planets that reached the pebble isolation mass and have accreted an atmosphere orbit close to the host star, with the exception of the model where the inner edge is placed further out for some planets. This shows that unless migration is halted, such as planets being trapped in resonances, planets with a significant atmosphere end up in orbits close to the host star because the migration timescales are significantly shorter than the gas accretion timescales. Planets injected close to the star grow quickly to the pebble isolation mass due to the high pebble flux close to the star. These planets experience efficient migration as the migration rate is proportional to the surface density of the gas \citep{tanaka2002_migration}. They therefore migrate to the inner edge rapidly. 

Comparing planets with similar core masses but different injection locations, we find that planets that were injected further away from star are able to accrete more gas compared to planets injected closer to the star if gas accretion is halted at the inner edge. While planets with similar core masses reach $\Miso$ at similar distances to the host star, planets that are injected further away from the star will generally reach $\Miso$ later on during the disc lifetime. This is a result of the gas surface density in the protoplanetary disc decreasing with distance from the star, causing the migration rates of planets to decrease. Planets injected further away will therefore migrate inwards slower due to their reduced migration rates. and have more time to accrete gas before reaching the inner edge of the disc. They can therefore reach a higher atmosphere mass fraction compared to planets with similar core masses that were injected closer to the star. 

As the luminosity of the star decreases over the disc lifetime, the temperature at a fixed location of the protoplanetary disc decreases over time. The luminosity of the planet increases with decreasing temperature at the outer boundary of the atmosphere (see Fig. \ref{fig:lum_sma}), which means that gas accretion can become more efficient later on during the disc lifetime. When gas accretion continues at the inner edge, we find that for planet with similar core masses, planets injected closer to the star are able to accrete more gas than planets injected further away from the star. This is contrasting the results when gas accretion is turned off at the inner edge. Planets injected closer to the star will reach the pebble isolation mass earlier compared to planets that were injected further away and will therefore be able to accrete gas for a longer time, reaching higher atmosphere mass fractions.

\subsection{Planet radii}
\label{ssec:pop_radii}
\begin{figure}
    \centering
    \includegraphics[width=\linewidth]{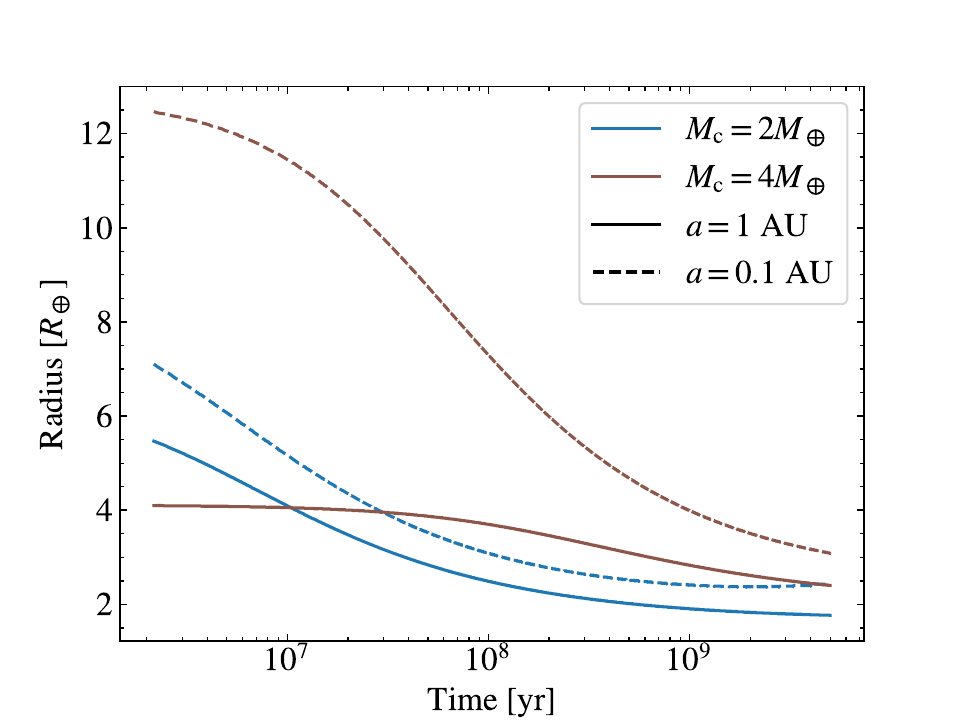}
    \caption{Radius as a function of time after disc dissipation for two different core masses and two different distances to the star. The planets have atmosphere mass fractions of 1\% and 5\% respectively, which are typical values for these core masses based on our simulations.}
    \label{fig:radius_time}
\end{figure}

In Fig. \ref{fig:radius_time}, we show the radius evolution as a function of time after disc dissipation for three different planets at two different distances to the star. The planets orbiting closer to star are initially more inflated due to their increased temperatures compared to the planets on wider orbits. It should be noted here that the initial radius is not equal to the outer integration boundary $R_{\rm out}$ as defined in Sect. \ref{ssec:gas_acc_lum}, which can reach up to 100 $R_\oplus$, but rather the radius as defined in Sect. \ref{ssec:contract_energy}, i.e. $\rrcb+\Delta R_{\rm tr}$ as defined in eq. \eqref{eq:R_tr}. Clearly, contraction is efficient and already after $\sim$1 Gyr, most of the contraction has already occurred and the sizes of the close-in planets have started to approach the planets on wider orbits. 

\begin{figure*}
    \centering
    \includegraphics[width=\linewidth]{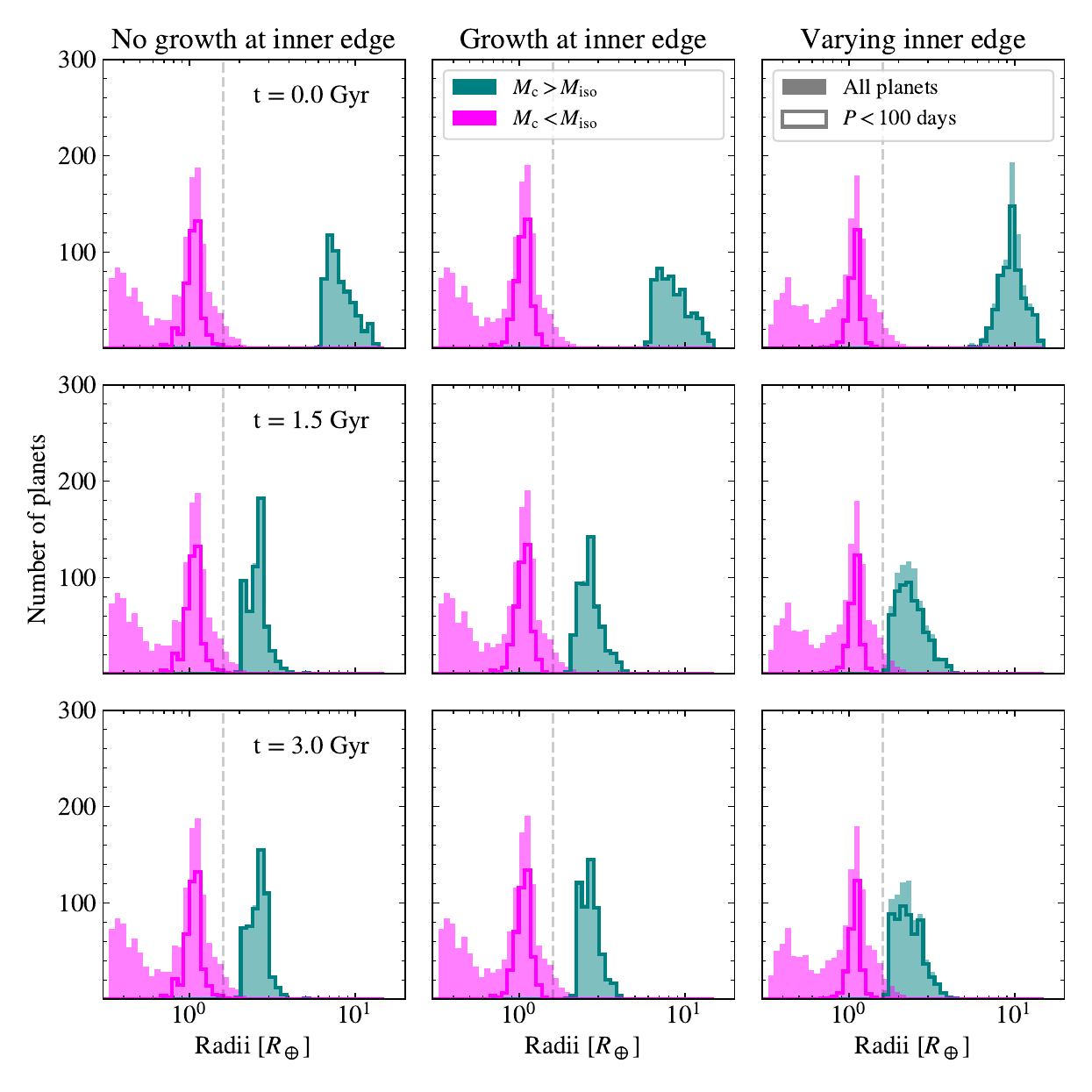}
    \caption{Histogram of the resulting planet radii for the three growth models at three different snapshots in time after disc dissipation. The filled histograms show the entire planet population while the unfilled histograms show planets with $P<100$ days. Clearly, the radius gap at $\sim$1.6 $R_\oplus$ (denoted by a dashed line) can exist without the need for mass loss. When the inner edge is at 0.1 AU, the outer peak is at $\sim$2.4-2.6 $R_\oplus$, in agreement with observations. However, when the inner edge varies between 0.1 AU and $\sim$1 AU, the radii of a large number of planets with atmospheres are significantly larger (3-4 $R_\oplus$) as they are more massive and therefore contract slower.}
    \label{fig:radii_hist}
\end{figure*}

In Fig. \ref{fig:radii_hist}, we show the radius distribution at three different snapshots in time for all three growth models where we have separated the planets with final masses above pebble isolation mass and planets below pebble isolation mass. The filled histogram shows the full population while the unfilled histograms only show planets with an orbital period less than 100 days. Clearly, all growth models show a final radius valley at $\sim$1.5-2 $R_\oplus$. Notably, the valley is significantly shallower in all models for the entire population compared to when we limit the data set to only planets with orbital periods less than 100 days. In the full population, the valley is mostly populated by bare planet cores with sizes of $\sim$2 $R_\oplus$. In order for a rocky core to reach such a large size, it would have to have a mass of $\sim$6 $M_\oplus$ while an icy core (with a water mass fraction of $\sim$20\%) can reach such a size with a mass of $\sim$4 $M_\oplus$. Should such a massive planet orbit close to the star, it would exceed the pebble isolation mass and therefore have accreted an atmosphere, inflating it beyond the radius valley.

Immediately after the dissipation of the disc, most planets that have reached the pebble isolation mass and accreted an atmosphere have inflated sizes larger than 10 $R_\oplus$. However, they are able to contract to $<$4 $R_\oplus$ already after 1.5 Gyr. The final radius distribution at 3 Gyr is similar to the distribution at 1.5 Gyr, which shows that the radius distribution is shaped relatively soon after disc dissipation, on $\sim$Gyr timescales. In the two models where the inner edge is fixed at 0.1 AU (left and middle column), contraction is slower as most of the planets orbit closer to the host star compared to the third model where the inner edge varies. When growth is halted at the inner edge, planets with atmospheres have radii around $\sim$1.5-2.5 $R_\oplus$, with a strong decline in the population for radii $>$2.4 $R_\oplus$, in agreement with observations \citep{fulton2017_valley}. The results when growth continues at the inner edge are similar, since the final atmosphere mass fraction distributions between these two models are similar. However, when the inner edge is allowed to vary between 0.1 AU and $\sim$1 AU, planets are able to accrete more massive atmospheres, resulting in some planets not being able to contract below $\sim$3-4 $R_\oplus$. We nevertheless find that the number of planets with sizes of $\sim$3-4 $R_\oplus$ is similar to the number of planets with sizes of $\sim$2-3 $R_\oplus$, which is not seen in the observed radius distribution. This could be a result of not including any mass loss processes as these highly inflated, low-mass planets would be more susceptible to mass loss processes. The location of the radius valley for close-in planets remain the same for all three models but due to efficient contraction further away from the star, the valley is more filled in when halting migration further out.

\begin{figure*}
    \centering
    \includegraphics[width=\linewidth]{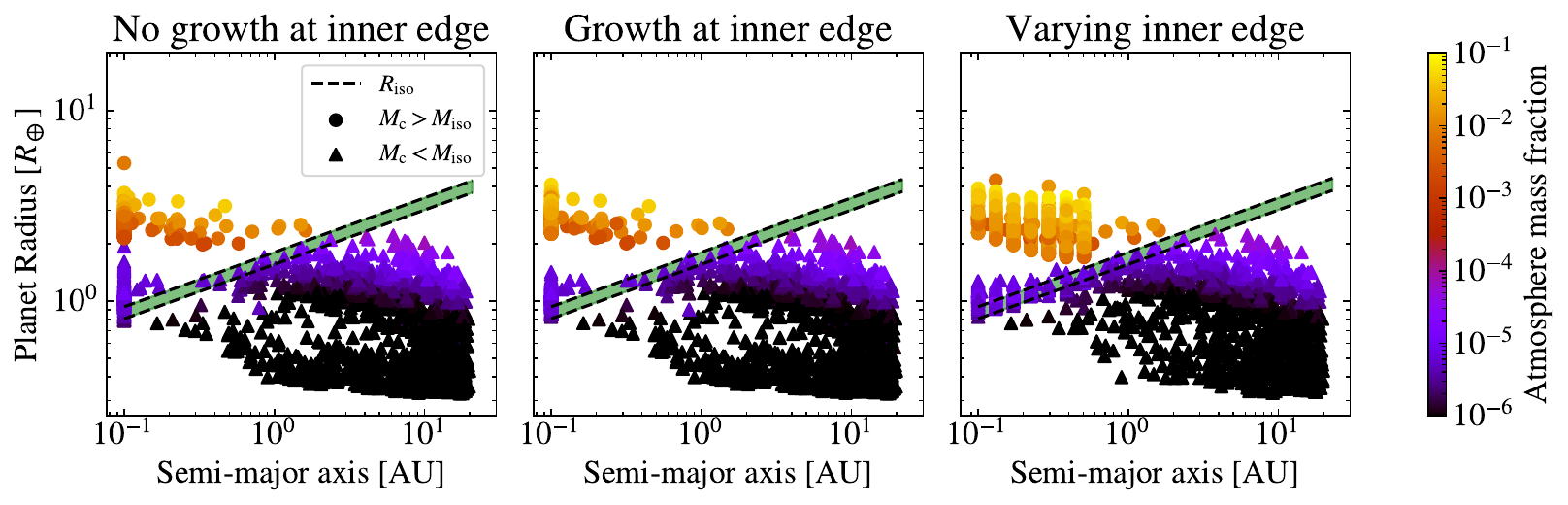}
    \caption{Planet radii as a function of their semi-major axis for all three models after 3 Gyr of evolution. The color of each point indicates the atmosphere mass fraction. The dashed lines show the core radius at the pebble isolation mass at the start and end of the disc lifetime for a planet with a density of 3.6 g/cm$^{3}$, which is a typical density for our planets. In all models, planets above pebble isolation mass that orbit further away from the star are smaller as their contraction rate is faster. When varying the inner edge, planets further away from the star are able to reach radii as low as $\sim$2 $R_\oplus$. However, planets that accrete a massive enough atmosphere ($\sim$10\% of core mass) remain large resist contraction.}
    \label{fig:radius_sma}
\end{figure*}

In Fig. \ref{fig:radius_sma} we show the radii of all planets after 3 Gyr as a function of their final semi-major axis. The region between the dashed lines is the radius for a planet at pebble isolation mass (and a density of 3.6 g/cm$^{3}$) throughout the disc lifetime. Some planets are larger than this radius without having accreted an atmosphere, which can be attributed to these planets either having a lower density or, as explained in Sect. \ref{ssec:pop_mass}, they have migrated to their final orbit towards the end of the protoplanetary disc life-time and therefore have not been able to accrete an atmosphere. In all models it is clear that planets further away from the star manage to contract more efficiently. When growth is allowed to continue at the inner edge (middle panel), some planets are able to accrete a slightly more massive atmosphere compared to when growth is halted at the inner edge (left panel). As a result, slightly fewer planets are unable to contract to sizes below 3 $R_\oplus$. When allowing the inner edge to vary, it is clear that planets with an inner edge outside of 0.1 AU manage to accrete significantly more massive atmospheres compared planet at 0.1 AU as a result of a faster contraction. A faster contraction after the dissipation of the protoplanetary disc results in a significant amount of planets contracting to $\sim$1.5-2 $R_\oplus$. Planets with atmosphere mass fractions $\gtrsim$a few percent resist contraction and remain large at 4-5 $R_\oplus$. Ultimately, the final radius distribution is significantly shaped by how close to the host star the planets end up orbiting. Planets orbiting further away from the star are able to accrete more gas but will inevitably contract faster as well. Further, the amount of gas that a planet can accrete is also determined by when, during the disc lifetime, it reaches the pebble isolation mass and whether contraction and subsequent gas accretion is allowed to continue at the inner edge. 

\subsection{Slope of the radius valley}
\label{ssec:valley_slope}

In order to estimate the slope of the radius valley as a function of orbital period, we want to fit a line to the radius gap in the planet population. We follow the procedure laid out in \citet{rogers2021_compare}. The radius gap can be defined as a function of the orbital period $P$.
\begin{equation}
\label{eq:gap_line}
    \log \rgap = m\log(P/P_0)+\log R_{\rm gap,0},
\end{equation}
where $P_0$ is a reference orbital period and $R_{\rm gap,0}$ is the position of the gap at $P_0$. In order to find the location of the gap, we calculate the probability density function (PDF) $f(\Rpl,P)$ of the planet population in $\log\Rpl-\log P$ space through a gaussian kernel density estimation. We then evaluate the integral
\begin{equation}
\label{eq:line_integral}
    \mathcal{I} = \int f(\log\Rpl,\log P)d\log P
\end{equation}
along the line defined in equation \eqref{eq:gap_line}. We can then find the gap line with parameters $m$, and $R_{\rm gap,0}$, which minimises $\mathcal{I}$. In order to estimate an uncertainty, we bootstrap the planet data with replacement and repeat the procedure 1000 times. We use an initial guess of $m=0$ and $\log (R_{\rm gap,0}/R_\oplus) = 0.2$ with a reference orbital period of $P_0=50$ days. We allow $m$ and $\log (R_{\rm gap,0}/R_\oplus)$ to vary in the intervals [-1,1] and [0,0.5] respectively. 

\begin{figure*}
    \centering
    \includegraphics[width=\linewidth]{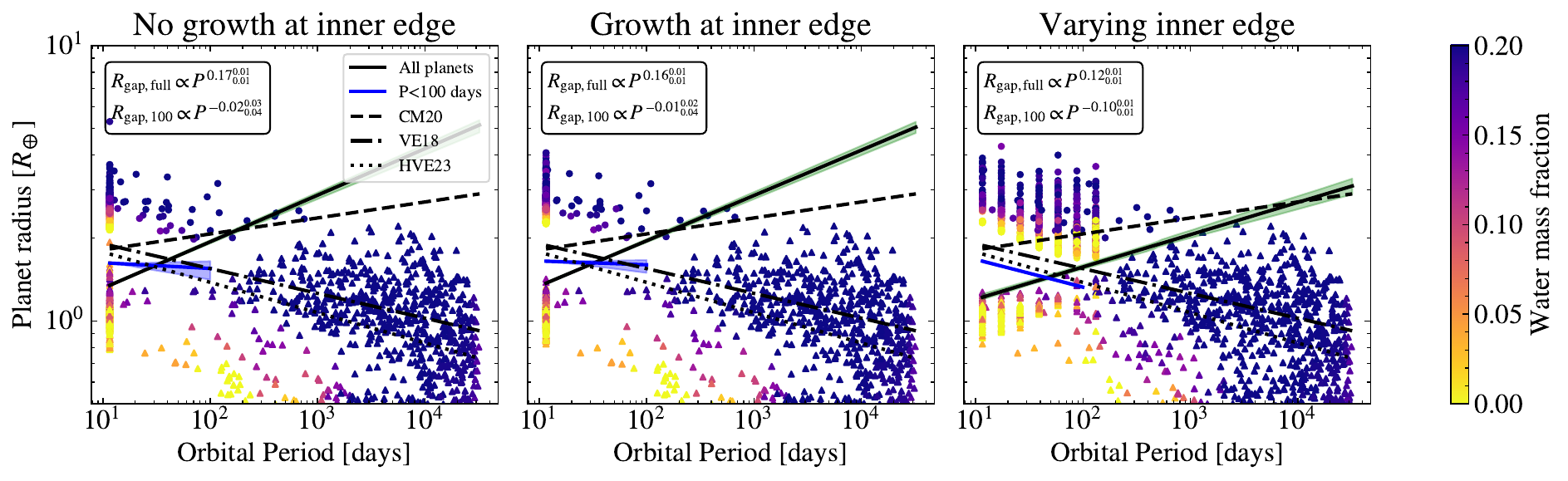}
    \caption{Planet radius as a function of orbital period for the three different models. Circles show planets above pebble isolation mass while triangles show planets below pebble isolation mass. The color of the points show the mass fraction of water. The solid lines show the fitted gaps in this work, both for the full population (black) and when limiting the population to only planets with orbital periods $<$100 days (blue). We also show the slopes from previous work by \citet{cloutiermenou2020_mdwarfslope} (CM20), \citet{vaneylen2018_gap} (VE18), and \citet{Ho_vaneylen2023_valleymstar} (HVE23). We also show the values for the fitted slopes.}
    \label{fig:gapline_period}
\end{figure*}

In Fig. \ref{fig:gapline_period}, we show the fitted gaps for the full planet population as well as only for planets with orbital periods $<100$ days. The color of each point denoted the mass fraction of condensed water of each planet. Triangles are planets below pebble isolation mass while circles show planets that have reached the pebble isolation mass and accreted a significant atmosphere. For the full population, the slopes of the gaps are positive with $\rgap$$\propto$$P^{0.17}$, $\rgap$$\propto$$P^{0.16}$, and $\rgap$$\propto$$P^{0.11}$ for the three models respectively. When limiting the data to planets with orbital periods $<$100 days, the direction of the slope shifts to become negative. The resulting radius gaps has negative slopes with $\rgap$$\propto$$P^{-0.03}$ and $\rgap$$\propto$$P^{-0.02}$, and $\rgap$$\propto$$P^{-0.11}$ respectively. In the two models with a fixed inner edge, the gap slopes are slightly shallower than derived slopes from observations \citep{vaneylen2018_gap,petigura2022_rgapvstar,Ho_vaneylen2023_valleymstar}. This discrepancy could be caused by e.g. observational biases, which we do not take into account here. The discrepancy could also be caused by our choice of gap fitting method, which could affect the final values of the slope \citep{berger2023_cpmobs}. In contrast, when the inner edge is varied, we find good agreement with the observed slope. The slope in this model is also close to the fitted slope by \citet{lee2022_gapwithoutmassloss}, although their slope is fitted to orbital periods up to 500 days, which might be a cause of the discrepancy.

We also note that when fitting the gap for the full population, we essentially recover the pebble isolation mass as the size of a bare planet core, with mass equal to the pebble isolation mass and a fixed density, will scale with orbital period as $P^{0.19}$. When limiting the data to close-in planets, the direction of the slope shifts as contraction is more efficient further away from the star, resulting in smaller planets sizes for increasing orbital periods for planets with a significant atmosphere.

By limiting to only planet with orbital periods $<$100 days, we remove most planets with sizes of $\sim$1.5-2 $R_\oplus$ in all three models. This is approximately the location of the radius valley. For an orbital period of 100 days around a solar mass star, $\Miso$$\approx$1-1.8 $M_\oplus$, resulting in a maximum core size of $\sim$1.4-1.7 $R_\oplus$ depending on the water mass fraction. This is roughly the location of the radius valley. Any planets exceeding these masses will have been able to cool down and start accreting gas, inflating their radii above the radius valley. The radius valley for close-in planets is therefore a natural consequence of the pebble isolation mass. It is however, also possible for planets to grow larger through mutual collisions after reaching the pebble isolation mass, potentially resulting in further growth as well as the loss of any accreted atmosphere, filling in the valley as well as populating the peak consisting of planets larger than the valley \citep{lambrechts2019_seform,izidoro2022_migvalley,pan2024_collisions}.

\subsection{Compositions of close-in planets}
\begin{figure*}
        \centering
    \includegraphics[width=\linewidth]{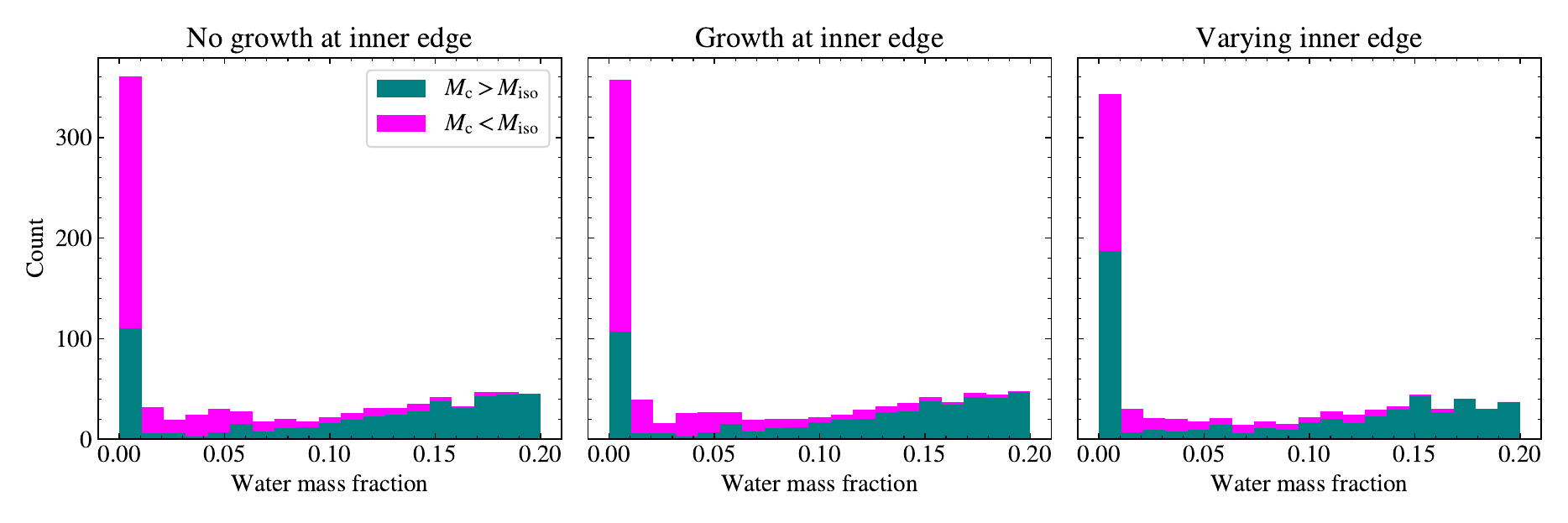}
    \caption{Mass fraction of accreted water for close-in planets ($P<100$ days). In magenta, we show planets with masses below the pebble isolation mass while in teal we show planets with masses above the pebble isolation mass. We find that in all models, planets with masses above the pebble isolation mass can be either rocky or water-rich while planets with masses below the pebble isolation mass are mostly water-poor.}
    \label{fig:wmf_dist}
\end{figure*}

In Fig. \ref{fig:wmf_dist}, we show histograms of the mass fractions of accreted water for all planets with orbital periods $<$100 days. A planet that has accreted all of its solid mass outside of the water iceline will have a water mass fraction of $\sim$20\% according to our chemical model used\footnote{See \citet{nielsen2023} for more details.}. We find, in all models, that the population of planets with masses above the pebble isolation mass consists of both rocky and water-rich planets. In contrast, planets below the pebble isolation mass are predominantly rocky, with some having accreted a small amount of icy pebbles. When varying the inner edge, the number of rocky planets with a significant atmosphere increases compared to the other two models as rocky planets injected inside of the water iceline can be trapped further out in the disc and accrete solids efficiently enough to reach the pebble isolation mass and grow beyond the radius valley. When the inner edge is set at 0.1 AU, planets injected inside of the water ice line quickly migrate to the inner edge where growth is either halted or limited, causing them to remain smaller than the radius valley. We therefore expect it to be rare to find water-rich planets that are smaller than the radius valley.

\section{Discussions and conclusions}
\label{sec:disc}
\subsection{Contraction compared to mass loss}
\label{ssec:disc_photoevap}
Our results indicate that under the assumptions presented in our models, mass loss is not necessary in order to create a radius valley. In fact, we find that atmosphere contraction alone can shape the radius valley on a $\sim$Gyr timescale, similar to the effect of core-powered mass loss. However, considering that XUV photoevaporation is expected to erode atmospheres on much shorter time scales (within a few 100 Myr), photoevaporation is expected to be more efficient at sculpting the shape of the radius valley compared to core-powered mass loss or models with no mass loss. Photoevaporation could cause compositional fractionation in the atmosphere as lighter species, such as hydrogen, would be more affected by the incoming XUV flux, enriching the upper atmosphere with heavier species. This would ultimately result in diffusion-limited photoevaporation as the upper atmosphere would need to be replenished by hydrogen through diffusion and can cause the mass loss rate to decrease significantly \citep{modirrousta_2024_difflimit}. Further, core-powered mass loss is expected to dominate over XUV photoevaporation for hot, low-density planets, while XUV photoevaporation dominates for denser planets further away from the star. Indeed, as the occurrence rate of sub-Neptunes decreases when comparing planets with expected ages of $\sim$Gyr with younger planets, observations seem to hint either towards photoevaporation being diffusion limited or core-powered mass loss being the dominant mass loss mechanism \citep{christiansen2023_k2ages}.

However, we stress that we do not claim that mass loss plays no role in shaping the radius distribution of close-in exoplanets. Indeed, some of the inflated planets with low core masses that orbit close to the star will most likely lose their primordial atmosphere as they are significantly hot and puffed up. When fitting the slope of the gap, we find best agreement using the model where we vary the inner edge. However, with this model, we find too many close-in planets with radii $\sim$3 $R_\oplus$ compared to observations, as can be seen in the lower right panel in Fig. \ref{fig:radii_hist}. This cannot be a result of detection bias as those planets would be easily detectable by the Kepler satellite. Therefore, it is expected that some mass loss processes, such as photoevaporation, also shape the radius distribution of close in planets. Should a significant fraction of the planets lose their atmosphere through some mass loss process, we expect that the population of planets smaller than the radius valley would get polluted by water-rich planets that lost their atmospheres. We also expect that the radius valley becomes shallower when mass loss processes are active given that a significant fraction of planets above pebble isolation mass have water-rich cores with sizes up to $\sim$1.5-2 $M_\oplus$. To demonstrate this, we show the final radius distribution when including photoevaporation for the migration model where we vary the inner edge in Appendix \ref{app:photoevap}. We find that several of the inflated planets lose their atmosphere after 3 Gyr and partially fill up parts of the radius valley. However, the peak at $\sim$2 $R_\oplus$ persists after photoevaporation. The excess of planets at $\sim$3 $R_\oplus$ could also be caused by the fact that we do not explicitly model the boil-off effect \citep{owenwu2016_boiloff}. After disc dispersal, the pressure support from the surrounding disc is gone, which can cause a rapid loss of mass from the now exposed planet atmosphere. This can, under certain conditions, result in as much as 90\% of the atmosphere to be lost \citep{rogers2024_boiloff}. The inclusion of the boil-off effect could therefore strip away a significant amount of gas from the planet, which would mimic the effects of photoevaporation. 

\subsection{Resulting semi-major axes}
\label{ssec:disc_sma}
Given that we let planets grow and evolve in isolation, we neglect the possible effects of multiple planets on the growth and migration of the planets. The formation of one planet could lock further planets in resonant chains, resulting in wider orbits \citep{peale1976_resonance,goldrecihtremaine1980_resonance,kajtazi2023_resonance}. Indeed, planets in young multi-planet systems have been observed to orbit just outside of mean-motion resonances indicating that planets can be prevented from migrate all the way in to the inner edge and instead get trapped in a resonance further away from the star \citep{fabrycky2014_resonances,hamer2024_mmrage}. These planets will be able to accrete more gas due to their higher luminosities compared to planets close to the star but will in turn contract faster after the dissipation of the protoplanetary disc. Therefore, in order to properly estimate the growth and subsequent contraction of planet atmospheres, it is clearly necessary to take into account possible effects which might prevent migration to the inner disc. Given that we evolve the planetary system over Gyr timescales, the inclusion of multiple planets could cause instabilities and, in some cases, even ejection of planets, further altering the architectures of the planetary systems \citep{mustill2017_instability}. The breaking of resonant chains could also result in giant impacts between planets that can strip away atmospheres and shape the radius valley further \citep{izidoro2021_breakingchains,izidoro2022_migvalley}. Recent work has also shown that it is possible to form massive planets cores and trap planets in e.g. pressure bumps, which would be an alternative mechanism for halting inwards migration of planets \citep{lau2022_pressurebump,sandor2024_pressurebump}. 

Further, \citet{gurrutxaga2023_outergiant} showed that if the pebble flux decreases significantly, planets can start accreting gas without reaching pebble isolation mass as the flux is low enough for the planets to start cooling down. This would result in planets accreting gas without having to grow to pebble isolation mass and thus not migrating inwards significantly, instead ending up on orbits further away from the star. Such a decrease could happen in the outer regions of the disc where the pebble reservoir is depleted first. However, in order for pebble drift to be efficient enough to deplete the pebble reservoir requires larger pebbles (St$\gtrsim$0.03) than we have in our work. 

\subsection{Effects of varying metallicity}
\label{ssec:disc_Z}
We have throughout this work assumed host stars with solar metallicity, which is generally true for stars in the Kepler catalogue \citep{berger2020a}. However, an increased metallicity could have significant effects on the final planet population. Most notably, the core mass distribution would shift to a more top-heavy distribution, which would affect the final radius distribution \citep{lee2022_gapwithoutmassloss,nielsen2023}. Furthermore, a higher metallicity would increase the opacity of the atmosphere and therefore decrease the luminosities of the planets, which would result in lower atmosphere mass fractions and less contraction of the planets after the dispersal of the disc. To illustrate the effect of opacity on the atmosphere contraction, we show the resulting radius distribution for our three simulations but with opacities of $0.1\kappa_{\rm BL}$ and $10\kappa_{\rm BL}$ in Appendix \ref{app:opacity_contraction}, representing a metal-poor and metal-rich atmosphere respectively. Clearly, increasing the opacity results in a reduced contraction rate with significantly inflated planets while reducing the opacity causes planets to contract significantly faster, although the effects of reducing the opacity are minimal compared to the nominal case as contraction is already efficient enough to contract the atmosphere considerably.
\subsection{The effects of water vapour}
\label{ssec:disc_water}
\begin{figure*}
    \centering
    \includegraphics[width=\linewidth]{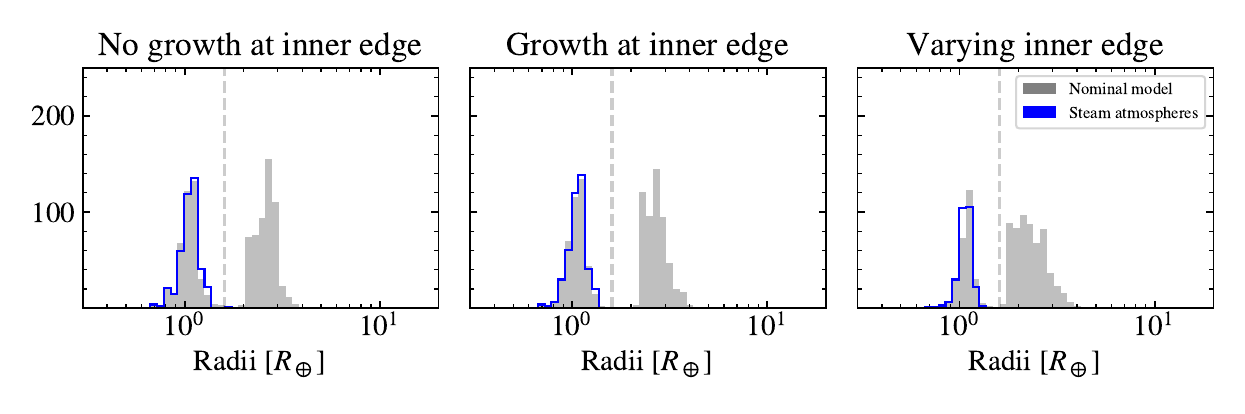}
    \caption{The radius distribution after 3 Gyr of evolution,  using a realistic interior structure that includes the compressibility of core and mantle material as well as steam atmospheres. The distributions should be compared to the lower panels of \ref{fig:radii_hist}. As our more detailed interior model compresses the cores compared to our simple core model, the increase in radii from the steam atmosphere results in the final radii not changing significantly compared to our nominal model.}
    \label{fig:radii_hist_water}
\end{figure*}

\begin{figure}
    \centering
    \includegraphics[width=\linewidth]{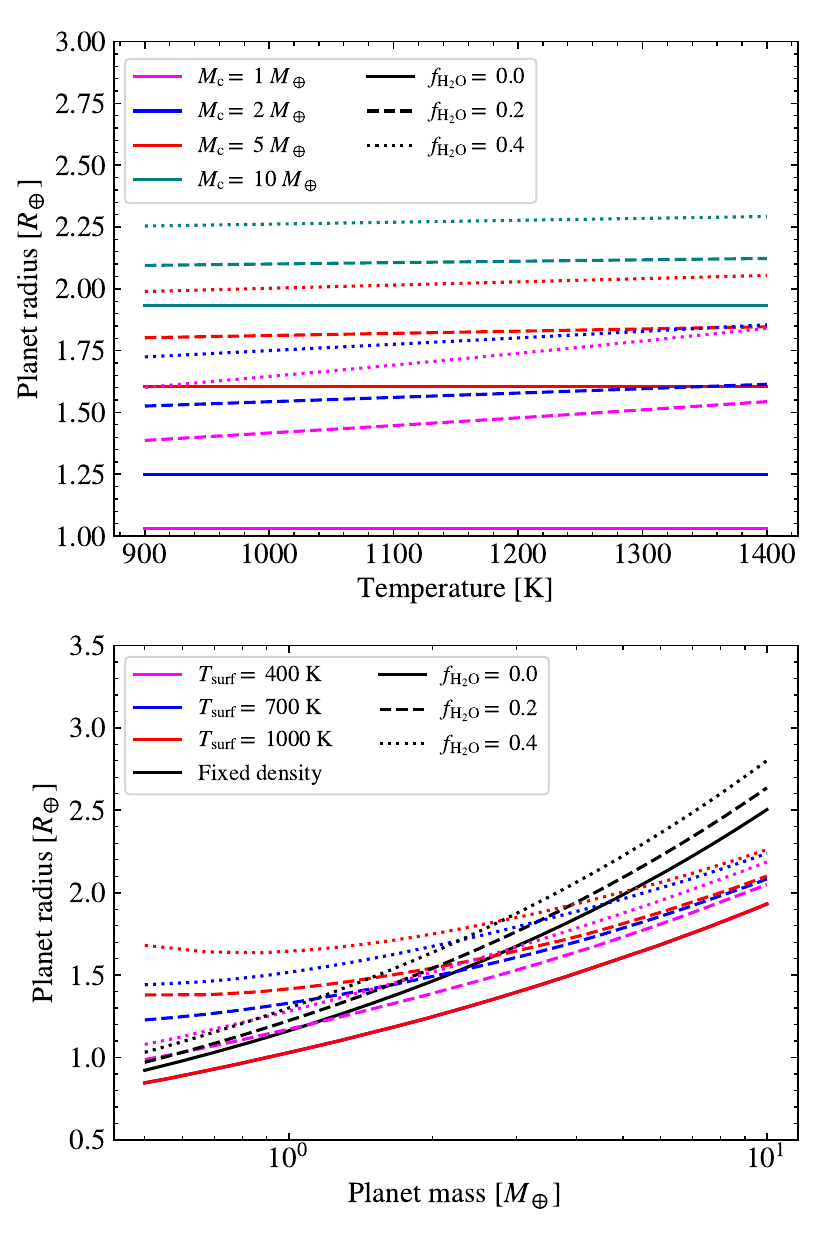}
    \caption{\textbf{Top:} Planet radius as a function of temperature for several planets with varying masses and water mass fractions. The effect of surface temperature is small in all cases, with the exception of the planet with a mass of 1 $M_\oplus$ and a high water mass fraction (40\%). Planets that occupy the radius valley or the sub-Neptune population requires either a high water mass fraction (40\%) or to be sufficiently massive ($>$2 $M_\oplus$ for water mass fractions of 20\% or $>$5 $M_\oplus$ for dry planets). \textbf{Bottom:} Planet radius as a function of planet mass for different surface temperatures. The different line types denotes the same water mass fractions as in the top panel. The black lines show our nominal radii using uncompressed densities of the core. The solid magenta and blue lines all overlap with the red solid line since the radius is independent of surface temperature when the water mass fraction is 0.}
    \label{fig:Rpl_Tsurf}
\end{figure}

Recently, works by \citet{venturini2020_water} and \citet{burn2024_waterworlds} found that the radius valley could be reproduced by considering that sub-Neptunes accreted a significant fraction of their mass as water ice outside of the water ice line and subsequently migrated inwards. By treating the accreted water as vapour mixed with the H/He atmospheres, they were able to reproduce the general shape and location of the radius valley. Throughout this work, we have assumed that the accreted atmospheres of the planets formed are purely H/He atmospheres. Therefore, we have only considered water in the shape of condensed ice, which means that it only contributes to the size of the planet cores. However, as seen in Fig. \ref{fig:gapline_period}, some planets have accreted some amount of water outside of the water ice line and have migrated inside the water ice line while remaining below the pebble isolation mass. This means that it might be necessary to treat their accreted water in the form of water vapour instead, using a more complex equation of state for their atmospheres. This would result in these planets being more inflated than expected through our model, potentially filling up parts of the radius valley. In order to test this effect, we included a model taking into account water in the form of steam and in a supercritical state. The details of the model are described in \ref{app:steam}. Input parameters for this model are: planet mass, water mass fraction, and surface temperature, which we take to be the irradiated temperature of the planet. From our chemical model, the highest possible water mass fraction of any planet is $\sim$20\%, as seen in Fig. \ref{fig:wmf_dist}. The resulting final radii are shown in Fig. \ref{fig:radii_hist_water}. Clearly, including steam atmospheres does not have a significant effect on the final results. This is partly caused by the fact that planets that are below the pebble isolation accreted pebbles mostly inside the water iceline and are therefore mostly water-poor (see Fig. \ref{fig:wmf_dist}). Further, the more complex interior model used to estimate the radii for our planets with steam atmospheres results in slightly smaller core sizes compared to our core radius calculations that were based on uncompressed densities. This has the effect of cancelling out the increase in radius due to the steam atmosphere.
The negligible change in radius could also be attributed to the fact that we have set the inner edge to be at 0.1 AU instead of closer to the star such as in \citet{liu2019_miso}, where the inner edge for a solar mass star is at 0.04 AU. At 0.1 AU, the irradiated temperature of the planet is $\sim$880 K while at 0.04 AU, it is $\sim$1400 K, which could result in a more inflated planet. In the top panel of Fig. \ref{fig:Rpl_Tsurf}, we show the planet radii for several planets as a function of temperature. Clearly, only the sizes of small planets with high water mass fractions are significantly inflated due to the change in temperature. Further, in order for planets with masses of$\sim$2 $M_\oplus$ to occupy the radius valley or the sub-Neptune population, planets need to have a high water mass fraction. For lower water mass fractions, planet masses need to be $\sim$5 $M_\oplus$ or above. In the lower panel of Fig. \ref{fig:Rpl_Tsurf}, we show the planet radius as a function of planet mass for different temperatures. We also show our nominal radii using uncompressed densities for different water mass fractions as black lines. Here, it is clear that we overestimate the radii in our nominal model as a dry Earth-mass planet is of similar size to an Earth-mass planet with a water mass fraction of 20\% at 400 K.
\subsection{Including more massive planet cores}
\label{ssec:massive_cores}
The most massive planet cores formed in our planet formation model have masses of $\sim$7 $M_\oplus$ due to our choice of pebble fragmentation velocity. However, the observed planet population include also massive planets with sizes $\sim$2-2.5 $R_\oplus$ and masses $\sim$10 $M_\oplus$ \citep[e.g.][]{zeng2019_MRobs}. These massive planets have also been formed in previous planet formation models \citep[e.g.][]{venturini2020_water}. Further, as seen in Fig. \ref{fig:Rpl_Tsurf}, dry planets with masses $\sim$10 $M_\oplus$ can be large enough to fill in the radius valley while more water-rich planets only need to grow to $\sim$2-5 $M_\oplus$ in order to fill in the radius valley. It is therefore of interest to investigate how the inclusion of more massive planet cores would affect this work. In Appendix \ref{app:v_frag}, we use fragmentation velocities of 1 m/s for dry pebbles inside the water iceline and 10 m/s outside the water iceline. As a result, the pebbles can grow larger in the outer disc, increasing the pebble flux and the accretion rate for protoplanets, which in turn results in some cores reaching 7-10 $M_\oplus$ as seen in Fig. \ref{fig:Mc_dist_vf}. We then evolve these planets for 3 Gyr, both with and without photoevaporative mass loss as described in Appendix \ref{app:photoevap}. The resulting radius distribution for planets with $P< 100$ days is shown in Fig. \ref{fig:radii_hist_varied_vfrag}. Further, we corrected the radii of planets below pebble isolation mass, taking into account the presence of steam atmospheres as described in Appendix \ref{app:steam}. Without mass loss but with the inclusion of more massive cores, the radius valley remains empty at the same location as in our nominal model. Instead, the number of planets with sizes $\gtrsim$ 3 $R_\oplus$ has increased as a result of the more massive planet cores. This is due to the fact that the pebble isolation mass at 100 days is $\sim$2 $M_\oplus$ meaning that all planets above this mass will have been able to accrete atmospheres up to a few percent of their mass, inflating them above the radius valley. When photoevaporative mass loss is included, we find that several of the planets previously larger than the radius valley lose some of or all of the atmosphere and shrink, partly filling up the radius valley as well as the peak below the radius valley. However, the location of the radius valley remains similar to our nominal model. From this, we can conclude that the radius valley is expected to consist of planets that have been partially or fully stripped of their atmospheres.

\subsection{Conclusions}
\label{ssec:conclusion}
In this work, we have simulated planet formation through pebble accretion and subsequent gas accretion after the pebble isolation mass is reached. We have then calculated the contraction of the accreted atmosphere after the disc has dissipated. From this, we conclude the following:
\begin{enumerate}
    \item We find that using our planet formation and evolution model a primordial radius valley is a natural consequence of some planets not accreting enough solid material to reach the pebble isolation mass. Only planets that reach the pebble isolation mass are able to accrete a significant atmosphere, getting inflated to radii above the valley. Despite this, we do not claim that mass loss is not a relevant physical process. Further, we investigate the effects of varying our chosen model parameters, such as the $\alpha$-viscosity, the emergence of a gap in the disc, the pebble fragmentation velocity, and an improved treatment of water in the mass-radius relationship. We find that the planet population varies little with respect to our choice of parameters, demonstrating that our conclusions are relatively robust to the choice of model and parameters.
    \\
    \item When including photoevaporative mass loss, the initially very empty radius valley becomes filled in by stripped cores with masses $>$2 $M_\oplus$. However, we find that the radius valley still persist with its location agreeing with observations. Further, we include a more careful treatment of water in our planet, allowing it to exist in the form of a steam atmosphere as well as partitioning it into the core and mantle. We find little to no difference in the resulting radii as the water mass fractions are relatively low ($<$20 \%). Additionally, our original estimate for radii of bare planet used uncompressed densities, which inflated their sizes compared to a more complex model. As a result, when including a more careful treatment of water, the increase in radii due to the steam atmosphere is cancelled out with the decrease in radii due to compression. Only when we add photoevaporation onto our complex water treatment do we find a significant difference in the resulting radius distribution with the radius valley being partially filled in.
    \\
    \item Migration plays a big role in shaping the final atmosphere mass fraction. If a planet with a low core mass is allowed to migrate freely, its atmospheric accretion will be limited, since it will quickly migrate close to the star where contraction and accretion is slow. In contrast, if migration is prevented, for example by trapping of the planets in a resonant chain, planets are able to accrete significantly more gas.
    \\
    \item For all planet migration models, we are able to reproduce the observed location of the radius valley for close in planets. When halting migration further out in the disc, we find that gas accretion becomes efficient enough to produce an excess of large planets with sizes of $\sim$3-4 $R_\oplus$ compared to observations as these atmospheres are massive enough to be resistant to significant contraction. Further, due to more efficient contraction further out in the disc of planets with low atmosphere mass fractions, the radius valley is more filled in when varying the inner edge compared to when the inner edge is fixed at 0.1 AU.
    \\
    \item We find that the radius valley gets filled in when extending to planets on wider orbits. Due to their high masses (4-5 $M_\oplus$) and water-rich composition, these planets are large enough to fill in the radius valley as they are not massive enough to reach the pebble isolation mass that far out in the disc. In contrast, planets with orbital periods $<$100 days will either be small enough to populate the peak on the small side of the radius valley or reach masses exceeding the pebble isolation mass, inflating them beyond the radius valley. 
    \\
    \item Finally, we find that in all models, the population of close-in planets larger than the radius valley consists of both rocky and water-rich cores. In contrast, close-in planets smaller than the radius valley are mostly water poor as they accreted pebbles inside of the water iceline and migrated inwards to the inner edge were their growth were halted and they remain bare, without a significant atmosphere. When varying the inner edge, a significantly higher number of rocky planets managed to reach the pebble isolation mass and grow beyond the radius valley as their migration was halted further out in the disc, allowing for efficient growth. 
\end{enumerate}
We believe that this work has contributed to our understanding of the properties and emergence of the radius valley. Our work importantly suggests that the radius valley is mostly present for close-in planets and that the valley is expected to be become gradually filled in when considering planets on wider orbits. We envision that future exoplanet detection missions such as PLAnetary Transits and Oscillations of stars (PLATO) \citep{plato} and the Nancy Grace Roman Telescope \citep{rst} will confirm this prediction by observing and characterising a significant number of planets with periods beyond 100 days. 
\begin{acknowledgements}
We thank the anonymous referee for their helpful comments, which have helped improve the quality of this work. A.J. acknowledges funding from the Danish National Research Foundation (DNRF Chair Grant DNRF159), the Carlsberg Foundation (Semper Ardens: Advance grant FIRSTATMO), the Knut and Alice Wallenberg Foundation (Wallenberg Scholar Grant 2019.0442) and the Göran Gustafsson Foundation. C.D. acknowledges support from the Swiss National Science Foundation under grant TMSGI2\_211313

\end{acknowledgements}
\bibliographystyle{aa}
\bibliography{bib}
\begin{appendix}
\section{Fragmentation velocity and the effect of core masses}
\label{app:v_frag}
\begin{figure}
    \centering
    \includegraphics[width=\linewidth]{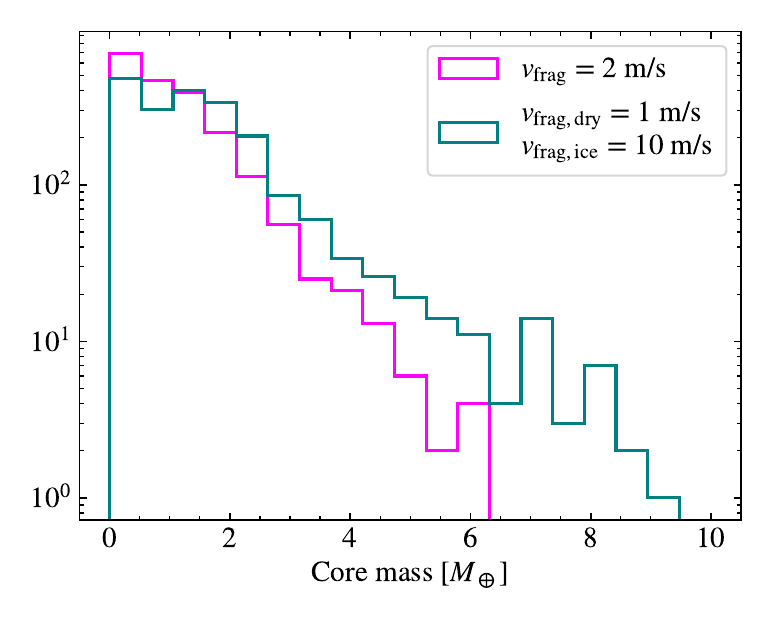}
    \caption{Core mass distribution for two different fragmentation velocity models. When varying the fragmentation velocity for water-rich and water-poor pebbles, more planets with core masses $\gtrsim$5 $M_\oplus$ are able to form compared to when using a flat fragmentation velocity of 2 m/s. The distribution of smaller cores is nevertheless relatively similar between the two models.}
    \label{fig:Mc_dist_vf}
\end{figure}

\begin{figure}
    \centering
    \includegraphics[width=\linewidth]{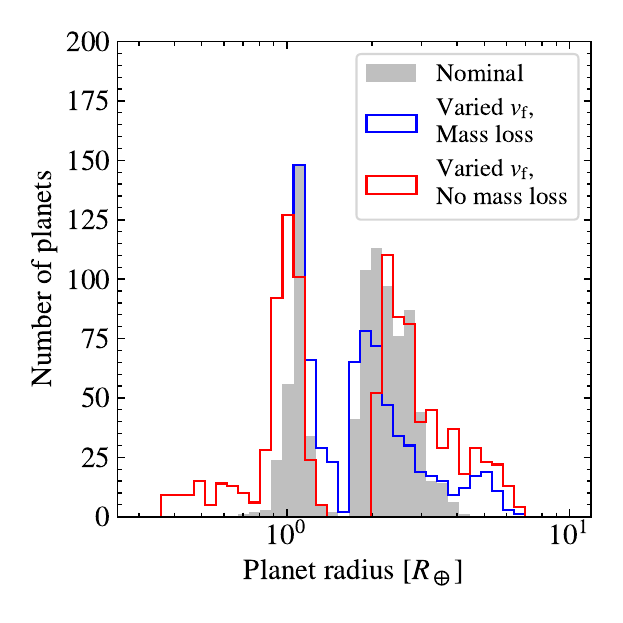}
    \caption{Radius distribution after 3 Gyr of evolution with a varying fragmentation velocity inside and outside the water ice line. The grey histogram shows the distribution using our nominal model without any mass loss while the blue histogram shows the same distribution but with  photoevaporative mass loss as described in Appendix \ref{app:photoevap}. We only consider our migration model where we vary the inner edge of the planets. Clearly, even when including more massive cores, we are able to reproduce the general location of the radius gap both with and without photoevaporative mass loss.}
    \label{fig:radii_hist_varied_vfrag}
\end{figure}

Throughout this work we adopted a fragmentation velocity of of pebbles of 2 m/s, motivated by our goal to reach St$\sim$0.01 in the outer regions of the disc. Previous work have instead adopted a fragmentation velocity of 1 m/s for dry pebbles inside the water ice line and 10 m/s for icy pebbles outside of the water ice line. In Fig. \ref{fig:Mc_dist_vf} we show the core mass distribution between these two models. When varying the fragmentation velocity, more planets with core masses $\gtrsim$5 $M_\oplus$ form compared to our adopted model. The number of planets with lower core masses nevertheless remains similar between the two models. As this study mainly focus on lower mass planets such as super-Earths and sub-Neptunes, we find that our adopted fragmentation velocity does not have a significant impact on our results. To further illustrate this, we show the final radius distribution after 3 Gyr for a planet population with a varying fragmentation velocity in Fig. \ref{fig:radii_hist_varied_vfrag}. We also include the distribution when photoevaporative mass loss is included as described in Appendix \ref{app:photoevap}. Further, in figure \ref{fig:radii_hist_varied_vfrag}, we also take into account the presence of steam atmospheres using our more complex interior model as described in Appendix \ref{app:steam}. Compared to our nominal model, the radius valley is partially filled in as more massive cores that have been stripped of their H/He atmospheres fill it in. In order for planets to be approximately the size of the radius valley using our maximum water mass fraction of 20\%, they need to have masses $>$2 $M_\oplus$ as we show in Fig. \ref{fig:Rpl_Tsurf}. Given that the pebble isolation mass at an orbital period of 100 days is 1-2 $M_\oplus$, planet that occupy the radius valley must therefore have been stripped of their primordial H/He atmosphere.
\section{Disc properties}
\label{app:disc_prop}
\begin{figure}
    \centering
    \includegraphics[width=\linewidth]{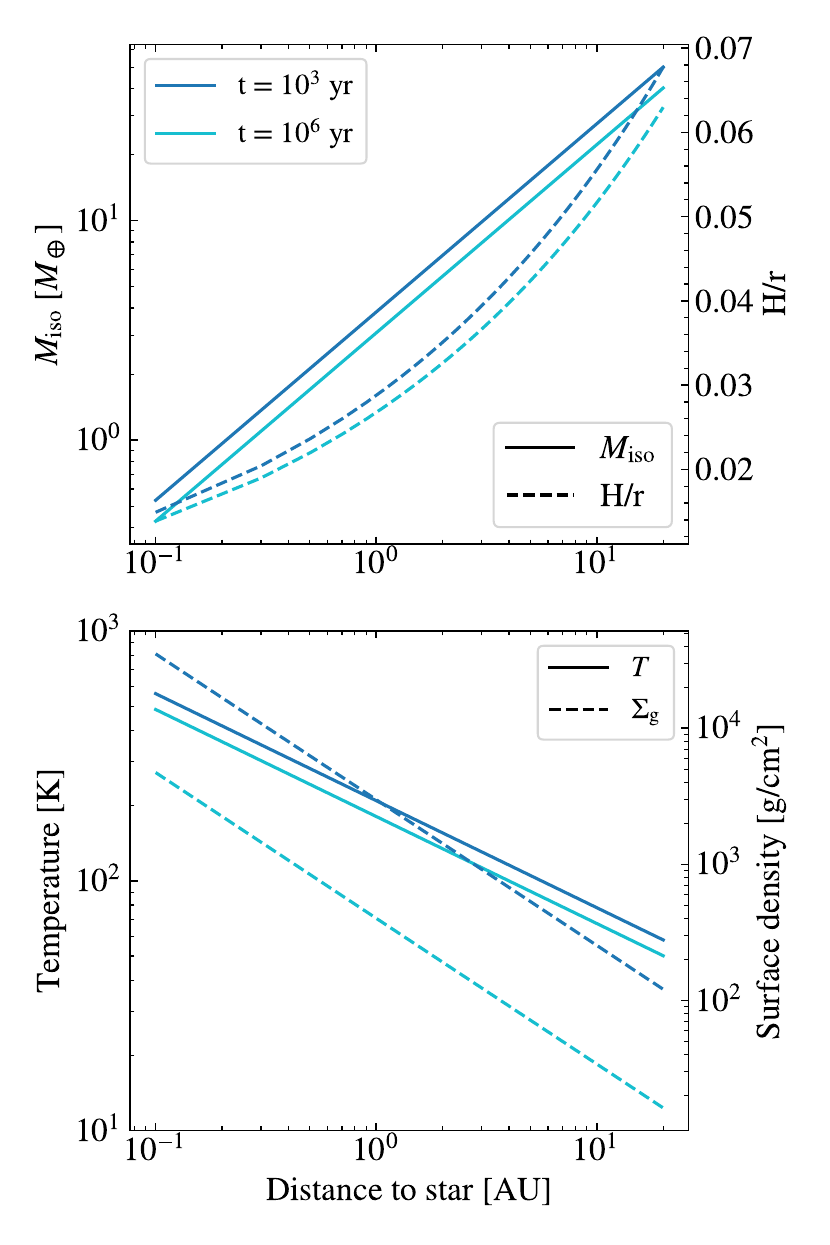}
    \caption{\textbf{Top:} Pebble isolation mass and H/r as a function of distance to the star at two different times. The aspect ratio H/r varies between 0.02 and 0.07, typical for an irradiated disc model. The pebble isolation mass varies from $\sim$0.5 $M_\oplus$ at 0.1 AU to $\sim$50 $M_\oplus$ at 20 AU. \textbf{Bottom:} Gas surface density and midplane temperature as a function of distance to the host star.}
    \label{fig:disc_prop}
\end{figure}

In Fig. \ref{fig:disc_prop} we show the pebble isolation mass and aspect ratio of the disc at two different times during the protoplanetary disc lifetime. We also show the initial gas surface density as well as the midplane temperature of the disc as a function of distance to the star in the bottom panel.
\section{Planet population for lower $\alpha$}
\label{app:lowalpha}
\begin{figure*}
    \centering
    \includegraphics[width=\linewidth]{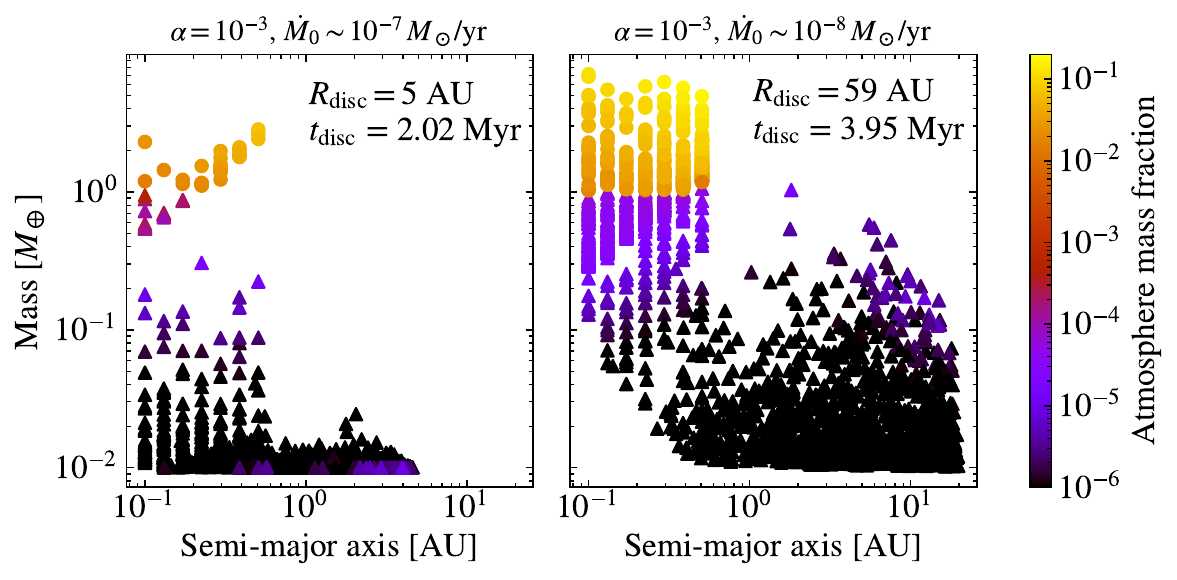}
    \caption{Similar to the rightmost panel in the top row in Fig. \ref{app:lowalpha}, but using $\alpha=10^{-3}$. In the left panel, we show the planet population with an initial gas flux of $\sim$$10^{-7}$ $M_\odot$/yr, which is the same as our nominal disc model. This results in a much smaller initial disc size, $R_{\rm disc}\sim$5 AU, while the disc lifetime remains comparable. We also show the case where we decrease the initial gas accretion rate with a factor of 10 in order to maintain the disc size. This results in a much longer disc lifetime and results in a relatively similar planet population as our nominal case.}
    \label{fig:low_alpha_pops}
\end{figure*}

As observations have shown that $\alpha$ could be as low as
$10^{-4}$-$10^{-3}$ \citep{trapman2020_alpha,rosotti2023_alphaobs}, we show the two planet populations, using $\alpha=10^{-3}$ in Fig. \ref{fig:low_alpha_pops}. We use the migration model where we vary the inner edge by assuming that the planet gets stuck in 3:2 resonances. The left panel shows the planet population using the same initial gas flux onto the star as in our nominal model, which results in a similar disc lifetime but a significantly smaller disc as the disc size is proportional to $\alpha$. For such a small disc, the gas surface density is significantly higher than in our nominal case, leading to an increased dust and surface pebble density and fast growth of planets. However, the dust in the disc is also lost to the star much quicker, resulting in a smaller window for planets to form. Further, as the rate of type-I migration is proportional to the surface density of the gas, migration is also more efficient, leading to the planets reaching the inner edge faster. As a result, the growth of planets is limited with most planets not being able to grow above 0.02 $M_\oplus$. The planet that manage to grow by being injected early on in the disc lifetime are quickly able to grow past the pebble isolation mass and start accreting a significant atmosphere.

In the right panel, we show the planet population but with a reduced initial gas flux onto the star in order to maintain the same disc size as in our nominal case. As the gas flux is now significantly lower, the disc lifetime increases to $\sim$4 Myr. As the disc size is similar to our nominal case, the result planet masses are similar to that of our nominal case. However, the longer disc lifetime results in more planets reaching high atmosphere mass fraction with the most massive planets reaching $\sim$20\%. Further, as migration is more efficient for a lower $\alpha$, we find fewer planets with masses $\gtrsim$1 $M_\oplus$ on wider orbits outside of $\sim$1 AU as most planets quickly migrate inwards and reach the pebble isolation mass, allowing them to accrete a more massive atmosphere.
\section{Gap formation in the protoplanetary disc}
\label{app:gap_disc}
\begin{figure}
    \centering
    \includegraphics[width=\linewidth]{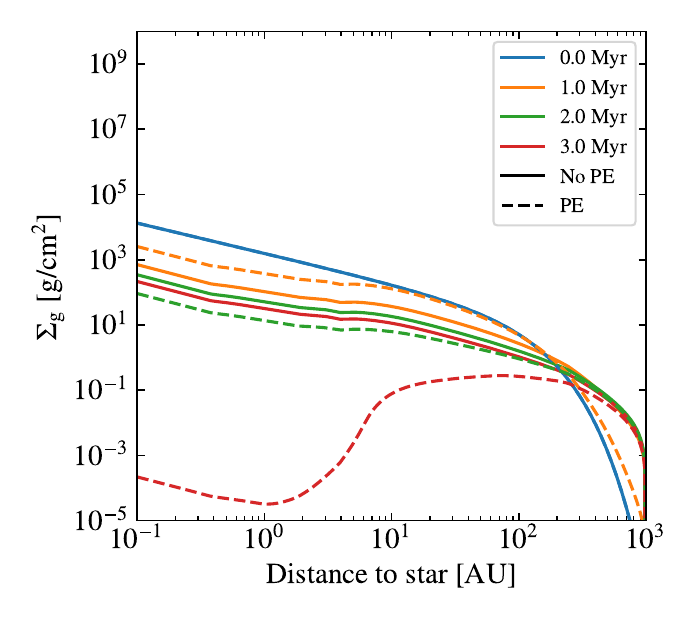}
    \caption{The evolution of the gas surface density using the viscous evolution in \texttt{chemcomp} for a few different snapshots in time. The gap opens up after $\sim$3 Myr, after which the disc is swiftly cleared. }
    \label{fig:disc_gap}
\end{figure}

\begin{figure}
    \centering
    \includegraphics[width=\linewidth]{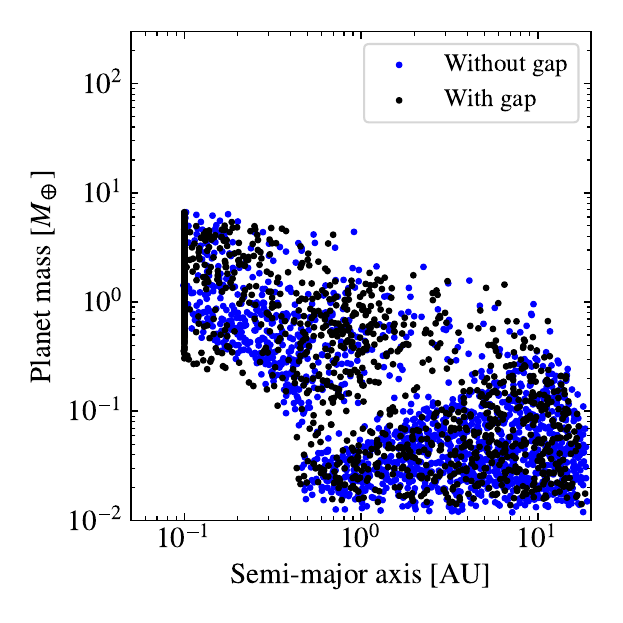}
    \caption{The resulting planet populations using our dust- and planet growth model but using the gas surface densities from Fig. \ref{fig:disc_gap}. The difference in the planet masses are insignificant, because the photoevaporative gap opens up too late in the life-time of the protoplanetary disc to significantly affect the migration and growth of the planets.}
    \label{fig:pop_comp_gap}
\end{figure}

In order to calculate the lifetimes for our protoplanetary disc, we adopted a simple photoevaporation model, which allowed us to calculate when all the mass from the disc was lost, either through accretion onto the star or through photoevaporation. However, the photoevaporation rate is not expected to happen uniformly throughout the disc, and as such, can create a gap in the protoplanetary disc. This gap could affect the growth of planet cores as dust and pebbles might fail to cross this gap. Further, migrating planets might get trapped at the gap location, preventing further migration. In our nominal model, we have implicitly assumed that the formation of a gap in the disc happens so late in the disc that it has little to no effect on the formation of the planets. As our analytical disc model is incapable of capturing the formation of a gap in the protoplanetary disc, we instead turn to a numerical method to test this assumption. We utilise the \texttt{chemcomp} code \citep{schneider2021_chemcomp}, which viscously evolves the disc by solving the viscous disc equation for a gas species Y\footnote{This is not the equation as presented in \citet{schneider2021_chemcomp} as they have included a source term $\Dot{\Sigma}_{\rm Y}$, which takes into account the sublimation of dust of a species Y as it crosses sublimation lines. This feature was turned off for our tests and therefore set to 0.}
\begin{equation}
    \frac{\partial \Sigma_{\rm g}}{\partial t} - \frac{3}{r}\frac{\partial}{\partial}\left[\sqrt{r}\frac{\partial}{\partial r}(\nu\sqrt{r}\Sigma_{\rm g})\right] = 0.
\end{equation}
We evolve the disc using as similar conditions as in our nominal model, with $\alpha=10^{-2}$, disc mass of 0.1 $M_\odot$, disc size of $\sim$60 AU but with a lifetime of 3.1 Myr instead of 2 Myr. For our photoevaporation model of the disc, we implemented the model by \citet{komaki2021_discPE} into \texttt{chemcomp}, which parameterise the change in the gas surface density as a result of X-rays, EUV, abd FUV radiation. The change is surface density can then be written as
\begin{equation}
    \log\left(\frac{\Dot{\Sigma}_{\rm g}}{1{\rm g\,cm}^{-2}\,{\rm s}^{-1}}\right) = c_5x^5+c_4x^4+c_3x^3+c_2x^2+c_1x+c_0,
\end{equation}
where $c_{5,...,0}$ are fit parameters, $x=\log(r/r_{\rm g})$, and $r_{\rm g}$ is the gravitational radius, given as 8.87 AU for a solar mass star. The fit parameters are given in table \ref{tab:PE_params} for a solar mass star. This model is only active in the range 0.1$r_{\rm g}\leq r \leq 20$$r_{\rm g}$. We show the resulting surface densities for a few different snpshots in time in Fig. \ref{fig:disc_gap}. Clearly, a gap opens up after $\sim$3 Myr.
\begin{table}[t]
    \centering
    \begin{tabular}{l|l}
     Parameter & Value \\\hline
     $c_5$ & 0.131 \\
     $c_4$ & -0.465 \\
     $c_3$ & 0.451 \\
     $c_2$ & 0.0.376 \\
     $c_1$ & -1.67 \\
     $c_0$ & -12.6 \\
    \end{tabular}
    \caption{Fit parameters in the \citet{komaki2021_discPE} model for photoevaporation of a protoplanetary disc around a solar mass star.}
    \label{tab:PE_params}
\end{table}
Given that \texttt{chemcomp} has a different planet growth model than our work, we tabulate the evolution of the surface density both with and without photoevaporation. We then grow 2000 protoplanets using our nominal dust- and planet growth model while reading off the surface density of the gas from the tabulated values at each time step. The resulting planet populations can be seen in Fig. \ref{fig:pop_comp_gap}. Both models yield very similar planet populations as the gap opens up very late during the disc lifetime, indicating that including the formation of a gap in the protoplanetary disc has a small effect on the resulting planet population.
\section{Energy of the core}
\label{app:E_c}
The specific thermal energy of the core is
\begin{equation}
    u_{\rm c} = \frac{k_{\rm B}}{\mu_{\rm c}m_{\rm H}}\frac{1}{\gamma_{\rm c}-1}T_{\rm c},
\end{equation}
where $\mu_{\rm c}$ and $\gamma_{\rm c}$ are the mean molecular weight and polytropic index of the core, respectively. Given typical temperatures at the bottom of the atmosphere ($\sim$$10^4$ K, see figure 1 in \citet{pisoyoudin}) the core is supercritical, convective and assumed to be nearly incompressible and therefore nearly isothermal. By inserting $T_{\rm c} = T(R_{\rm c})$ from \eqref{eq:T_prof}, we get 
\begin{equation}
\begin{split}
    & u_{\rm c} = \frac{\trcb k_{\rm b}}{\mu_{\rm c}m_{\rm H}(\gamma_{\rm c}-1)}\\
    & +\frac{\mu}{\mu_{\rm c}}\frac{1}{\gamma}\frac{\gamma-1}{\gamma_{\rm c}-1}\frac{G\Mpl}{\Rc}-\frac{\mu}{\mu_{\rm c}}\frac{1}{\gamma}\frac{\gamma-1}{\gamma_{\rm c}-1}\frac{G\Mpl}{\rrcb} \\
    & = \frac{\trcb k_{\rm b}}{\mu_{\rm c}m_{\rm H}(\gamma_{\rm c}-1)}+G\Mpl\frac{\mu}{\mu_{\rm c}}\frac{1}{\gamma}\frac{\gamma-1}{\gamma_{\rm c}-1}\left(\frac{1}{\Rc}-\frac{1}{\rrcb}\right).
\end{split}
\end{equation}
The total thermal energy of the core is then just $u_{\rm c}M_{\rm c}$. Clearly, when $\rrcb\gg\Rc$, this term is constant with respect to $\rrcb$ and can therefore be omitted in our contraction model.
\section{Including photoevaporation}
\label{app:photoevap}
\begin{figure}
    \centering
    \includegraphics[width=\linewidth]{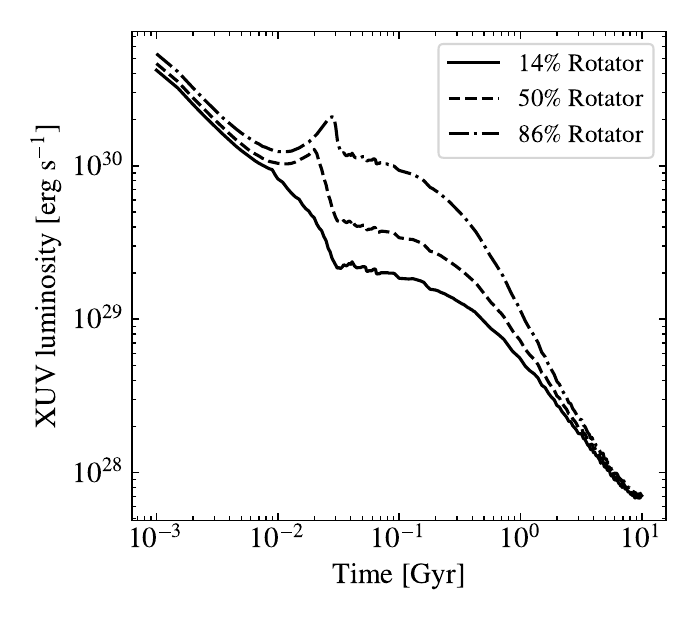}
    \caption{XUV-fluxes for stars with rotation rates equal to different percentiles of the total distribution at 150 Myr, from \citet{johnstone2021_xuv}. For example, the 50\% rotator then has a rotation rate equal to the 50 percentile at an age of 150 Myr.}
    \label{fig:xuv_fluxes}
\end{figure}

\begin{figure}
    \centering
    \includegraphics[width=0.9\linewidth]{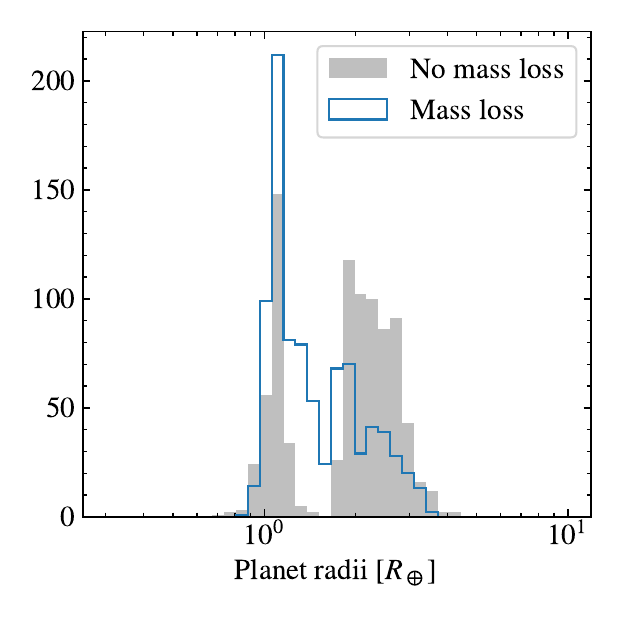}
    \caption{Radius distribution after 3 Gyr when taking into account photoevaporative mass loss. We only show the results for the migration model where we vary the inner edge. Further, for planets below the pebble isolation mass or planets where the atmosphere was completely stripped, we recalculate the radii by including a steam atmosphere as described in Appendix \ref{app:steam}. Several of the planets with sizes $>$2 $R_\oplus$ lost a significant part of their atmosphere, filling up the radius valley slightly. However, the valley persists even after photoevaporation.}
    \label{fig:radii_hist_photoevap}
\end{figure}

Given that some of the highly inflated planets that form in our models are expected to lose their atmospheres through e.g. photoevaporation, it is likely that the radius valley in our model will become partially filled in by more massive cores hosting light H/He atmospheres that were lost due to photoevaporation. We therefore test the effects of X-ray photoevaporation on the final radius distribution. During the protoplanetary disc lifetime, we used the simple relation from \citet{bae2013_xray} to calculate the photoevaporative loss of the gas disc. However, in order to fully capture the effects of photoevaporation over Gyr timescales, we instead turn to the work by \citet{johnstone2021_xuv}, which includes tabulated values for the X-ray luminosity for different stellar masses and rotation rates. The input rotation rates given in the tables of \citet{johnstone2021_xuv} can either be the initial rotation rate or the percentile of a distribution of rotation rates at an age of 150 Myr. For our photoevaporation model, we therefore choose the flux from a star with rotation equal to the 50th percentile, which we choose to denote ''50\% rotator''. We show the temporal evolution of XUV-fluxes for different percentiles in Fig. \ref{fig:xuv_fluxes}.
We include a simple prescription of photoevaporation rates of the planet given by \citet{owen_wu2017_PE} where the mass-loss from photoevaporation can be given as 
\begin{equation}
    \Dot{M}_{\rm PE} = \eta\frac{\Rpl^3 L_{\rm XUV}}{4a^2G\Mpl},
\end{equation}
where $\eta$ is an efficiency factor that captures the efficiency of atmosphere removal. Similarly to \citet{owen_wu2017_PE}, we adopt $\eta=0.1$. We show the final radius distribution after 3 Gyr using the model where we vary the inner edge of planets in Fig. \ref{fig:radii_hist_photoevap}. We also correct the radii for planets with masses below pebble isolation mass as well as completely stripped planets, taking into account the fact that any water present would exists as a steam atmosphere. We describe this correction in Appendix \ref{app:steam}. When comparing to our nominal model without any mass loss, several of the inflated planets in our nominal model lost most of their atmospheres and shrunk, filling up parts of the radius valley. However, the radius valley persists. We note that the second peak lies at $\sim$2 $R_\oplus$, which is close to the observed peak, which lies at $\sim$2.1 $R_\oplus$ \citep{fultonpetigura2018_gap}. We find however that there is a steep drop-off after $\sim$2 $R_\oplus$ compared to the observed drop-off at 2.4 $R_\oplus$. This could be caused by the fact that our model lacks planets with masses $>$7 $M_\oplus$. These planets would be mostly resistant to photoevaporation and would therefore fill up the tail end of the distribution. It is also possible that including the boil-off effect could cause planets with sizes $\gtrsim$3 $R_\oplus$ to shrink further and shrink to 
$\sim$2-2.5 $R_\oplus$. However, we would like to stress that we are not aiming to fit the observed radius distribution but rather try to understand whether the radius valley is primordial or a result from mass-loss processes. The inclusions of these larger planets or the boil-off effect would therefore not have a significant effect on the conclusions of this work.
\section{The effect of atmosphere opacity on contraction rate}
\begin{figure*}
    \centering
    \includegraphics[width=\linewidth]{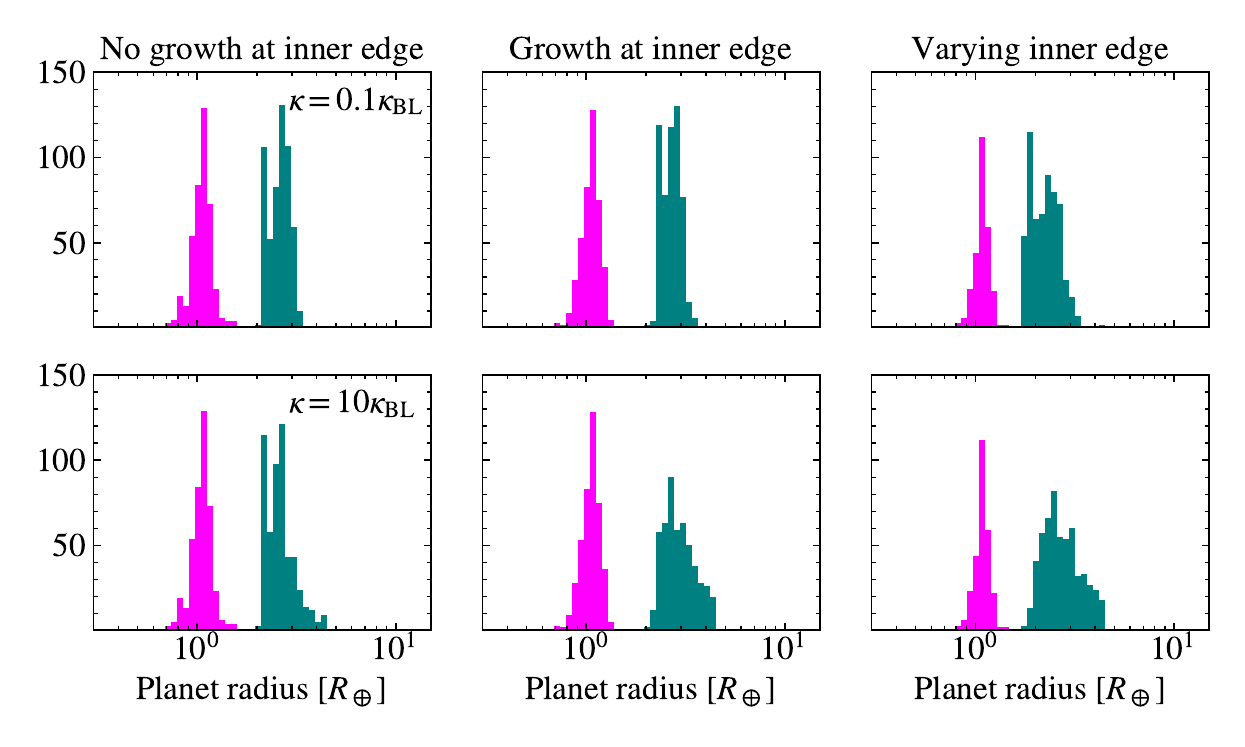}
    \caption{The resulting radius distribution for all of our models for two different opacities after 3 Gyr evolution. The top row shows the resulting distribution after contraction with $\kappa = 0.1\kappa_{\rm BL}$ while the bottom row shows the resulting distribution with $\kappa = 10\kappa_{\rm BL}$. Decreasing the opacity with a factor of 0.1 does not affect the final distribution significantly as contraction is already efficient in our nominal case. In contrast, increasing the opacity significantly slows down contraction such that a significant fraction of planets remain large at $\sim$3-4 $R_\oplus$.}
    \label{fig:radii_hist_opacity}
\end{figure*}

The opacity in the atmosphere  determines how fast energy can be radiated away and therefore how quickly the atmosphere contracts. In the main text, we used the opacity law from \citet{bell_lin1994_opacities}. In Fig. \ref{fig:radii_hist_opacity} we show the resulting radius distribution using the same planet population from the three models in the main text but where we modify the atmosphere opacity during contraction by a factor of 0.1 and 10 respectively. Clearly, for a low opacity, contraction is significantly more efficient compared to when the opacity is increased. For lower opacities, the maximum size in all models is $\sim$3 $R_\oplus$ while in the case with increased opacities, the maximum planet size is $\gtrsim$4 $R_\oplus$. 
\label{app:opacity_contraction}
\section{Steam atmosphere model}
\label{app:steam}
As previously shown by \citet{venturini2020_water} and \citet{burn2024_waterworlds}, allowing water to exist in the form of vapour as a steam atmosphere can lead to larger radii of planets with small or absent H/He atmospheres. This effect could cause several of the smaller planets in this work to become inflated and fill in the radius valley. In order to test the effect of water in the planet radii, we need a model for planets below the pebble isolation mass (i.e. without a significant H/He atmosphere) but with some amount of accreted water. We employ an advanced interior model that is based on \citet{dorn_generalized_2017} with recent adaptations as in \citet{luo_majority_2024}. For this application, we focus on interiors with Earth-like rocky interiors without H/He atmospheres and only pure steam atmospheres. Water can be present in the core, the mantle, and at the surface, depending on the specific thermal state of the planets.

We consider a core made of Fe with the light alloy elements H and O. For solid Fe, we use the equations of state for hexagonal close packed (hcp) iron \citep{hakim_new_2018}. For liquid iron and the iron alloys, we use \citet{luo_majority_2024}. The core thermal profile is assumed to be adiabatic throughout the core. At the core-mantle boundary (CMB), there is a temperature jump as the core can be hotter than the mantle due to the residual heat released during core formation following \citet{stixrude_melting_2014}.

The mantle is assumed to be made up of three major constituents, i.e., MgO, SiO$_2$, FeO. Here, we assume Earth-like composition. For the solid mantle, we use the thermodynamical model \texttt{Perple}$_X$ \citet{connolly_perplex}, to compute stable mineralogy and density for a given composition, pressure, and temperature, employing the database of \citet{stixrude_thermal_2022}. For pressures higher than $\sim$125 GPa, we define stable minerals a priori and use their respective equation of states from various sources \citep{fischer_equation_2011, faik_equation_2018, hemley_constraints_1992, musella_physical_2019}. For the liquid mantle, we calculate its density assuming an ideal mixture of main components (Mg$_2$SiO$_4$,SiO$_2$,FeO) and add them using the additive volume law. Note that we use Mg$_2$SiO$_4$ instead of MgO since the data for forsterite has been recently updated for the high-pressure temperature regime, which is not available for MgO to our knowledge. The mantle is assumed to be fully adiabatic.

Water can be present in mantle melts, while the solid mantle is assumed to be dry. The addition of water reduces the density, for which we follow \citet{bajgain_structure_2015}. For small water mass fractions, this reduction is nearly independent of pressure and temperature. The melting curve of mantle material is calculated for dry and pure MgSiO$_3$ to which the addition of water \citep{katz_new_2003} and iron \citep{dorn_interior_2018} can lower the melting temperatures. Water that is added to the core, will also lower its melting temperature, for which we follow \citep{luo_majority_2024}. Water that is present in the core, can be present in both liquid and solid phase. The partitioning between mantle melts and the water layer is determined by Henry’s law, for which we use the fitted solubility function of \citep{dorn_hidden_2021}. For the partitioning of water between iron and silicates, we follow \citet{luo_majority_2024}. For the equilibration pressure of water to partition between iron and silicates, we use half of the core-mantle boundary pressure. For the planets of interest, water is in steam or supercritical phase, for which we use the EoS compilation AQUA \citep{haldemann_aqua_2020}. For pressures below 0.1 bar, we assume an isothermal profile and switch to an adiabatic profile.
\end{appendix}
\end{document}